\documentclass[a4paper,10pt,twocolumn]{article}
\pdfoutput=1

\usepackage[utf8]{inputenc}
\usepackage[british]{babel}
\usepackage[T1]{fontenc}
\usepackage{lmodern}
\usepackage[margin=2cm,columnsep=0.8cm]{geometry}
\usepackage{amsmath,amssymb}
\usepackage{graphicx}
\usepackage{authblk}
\usepackage[numbers]{natbib}
\usepackage[colorlinks,allcolors=blue]{hyperref}

\graphicspath{{imagenes/}{./}}

\hypersetup{
  pdftitle={Modelling Josephson Junctions with Boundary Conditions: Towards High-Fidelity Models for Superconducting Qubits},
  pdfauthor={Luis Bueno Blanco, Juan Manuel P\'erez-Pardo},
  pdfkeywords={Josephson junctions, superconducting qubits, self-adjoint extensions, boundary conditions, current-phase relation, higher Josephson harmonics}
}

\title{Modelling Josephson Junctions with Boundary Conditions: Towards High-Fidelity Models for Superconducting Qubits}

\author[1]{Luis Bueno Blanco}
\author[1]{Juan Manuel P\'erez-Pardo\textsuperscript{*}}
\affil[1]{Departamento de Matem\'aticas, Universidad Carlos III de Madrid,
Avda.\ de la Universidad 30, 28911 Legan\'es, Madrid, Spain}
\date{}

\begin{document}

\twocolumn[
\begin{@twocolumnfalse}
\maketitle
\begin{abstract}
Josephson junctions are the central nonlinear element of superconducting qubits, yet
their standard description is reached through a chain of approximations.
We propose a
purely quantum-mechanical model of the junction, formulated directly in Hilbert space:
a charged particle moves on a one-dimensional loop, and the junction is a point at which
the wave function obeys self-adjoint boundary conditions, 
labelled
by a coupling strength $\lambda$, which controls the transparency of the junction, and a
phase jump $\delta$, which plays the role of the superconducting phase difference.
For a loop with a single junction we solve the spectral problem and show that
the ground-state momentum expectation, identified with the supercurrent, factorises
exactly as $\langle\hat{p}_0\rangle = I_c(\delta,\lambda)\sin\delta$. For opaque junctions
$I_c$ becomes independent of $\delta$ and the first Josephson relation $I=I_c\sin\delta$
is recovered; for strongly attractive coupling, where the ground state is bound to the
junction, the $\sin\delta$ law is also recovered, but with an exponentially suppressed
amplitude. For more transparent junctions, however, the model \emph{predicts} deviations
from the pure $\sin\delta$ law: $I_c$ acquires a dependence on $\delta$ already at order
$\lambda^{-2}$, compatible with the higher Josephson harmonics recently found necessary to
describe superconducting qubit spectra, and a fully transparent contact yields a current
linear in $\delta$. When the phase varies in time, a gauge transformation
shows that the voltage drop across the junction is proportional to $\dot\delta$, so
that the second Josephson relation emerges without being imposed. The boundary-condition
approach thus reproduces both Josephson relations from first principles and predicts
corrections to them, offering a route towards higher-fidelity models of superconducting
qubits.
\end{abstract}

\medskip
\noindent{\small\textit{Keywords:} Josephson junctions, superconducting qubits,
self-adjoint extensions, boundary conditions, current-phase relation, higher
Josephson harmonics.\par}

\smallskip
\noindent{\small\textit{Mathematics Subject Classification (2020):} 81Q10, 81Q35, 82D55, 81P68.\par}
\bigskip
\end{@twocolumnfalse}
]
{\renewcommand\thefootnote{*}%
\footnotetext{ORCID: \href{https://orcid.org/0000-0003-2424-8943}{0000-0003-2424-8943}}}

\section{Introduction}
\label{sec:introduction}


\subsection{Superconducting Qubits and Josephson Junctions}

The pursuit of scalable quantum computing has established superconducting
circuits as one of the leading experimental platforms for quantum information
processing~\cite{devoret2013}. Among the various physical realisations of a
qubit---the elementary unit of quantum information---superconducting qubits stand
out for their tunability, lithographic fabricability, and steadily improving
coherence times, which have grown by more than five orders of magnitude~\cite{jiang2025,bland2025} since the
first demonstration in 1999~\cite{nakamura1999}. Current multi-qubit processors
routinely integrate more than a hundred superconducting qubits on a single
chip~\cite{kim2023,google2025,gao2025}, and
their design, characterisation, and control constitute a vibrant area of both
physics and engineering.

At the heart of every superconducting qubit lies a \emph{Josephson junction}: a
thin insulating barrier (typically a few nanometres of aluminium oxide)
separating two superconducting electrodes. The junction is the only known
electronic circuit element that is simultaneously nonlinear and essentially
non-dissipative at millikelvin temperatures, and it is this combination that
makes superconducting qubits possible.

The fundamental physics of the junction was predicted by Josephson in
1962~\cite{josephson1962} and subsequently confirmed experimentally. Two
relations govern the dynamics. The \emph{first Josephson relation} expresses the
supercurrent $I$ through the junction in terms of the phase difference $\delta$
between the macroscopic wave functions of the two superconductors:
\begin{equation}
I = I_c \sin\delta,
\label{eq:josephson-current}
\end{equation}
where $I_c$ is the critical current, the maximum supercurrent the junction can
sustain without developing a voltage. The \emph{second Josephson relation}
connects the voltage $V$ across the junction to the time derivative of the phase:
\begin{equation}
V = \frac{\hbar}{2e}\,\frac{\mathrm{d}\delta}{\mathrm{d}t}.
\label{eq:josephson-voltage}
\end{equation}

Together, these relations endow the junction with a nonlinear, non-dissipative
inductance $L_J(\delta) = \hbar/(2eI_c\cos\delta)$ and a periodic potential energy
\begin{equation}
U(\delta) = -E_J\cos\delta, \qquad E_J := \frac{\hbar\,I_c}{2e},
\label{eq:josephson-energy}
\end{equation}
where $E_J$ is the Josephson energy. 

In the circuit-QED architecture~\cite{blais2021}, a Josephson junction is
embedded in a superconducting resonator or loop. The effective Hamiltonian of the
resulting qubit is not obtained directly from a microscopic theory of the
junction. Instead, the standard description follows a hierarchical procedure that
is systematically reviewed in~\cite{wendin2007,wendin2017}. 
As a starting point one first invokes the BCS theory of superconductivity, 
following Josephson's original approach~\cite{josephson1962}, 
to obtain the constitutive macroscopic
relations of the junction, Eqs.\ \eqref{eq:josephson-current} and \eqref{eq:josephson-voltage}.
These
macroscopic laws are then incorporated, together with Kirchhoff's rules, into a
lumped-element \emph{classical} electrical circuit, whose dynamics is formulated
using classical Hamiltonian mechanics. Several pairs of canonically conjugate
variables are possible, e.g.\ charge and phase, which lead to different classical Hamiltonians
describing the classical circuit~\cite{wendin2007}. In the next step,
this classical circuit model is \emph{quantised}: the conjugate variables, such as
the phase difference and the charge (or Cooper-pair number), are promoted to
operators satisfying canonical commutation relations. One arrives at a quantum 
Hamiltonian, for instance: 
\begin{equation}
H = 4E_C\,(\hat{n} - n_g)^2 - E_J\cos\hat{\delta},
\label{eq:qubit-hamiltonian}
\end{equation}
where $E_C = e^2/(2C)$ is the charging energy, $\hat{n}$ is the Cooper-pair number
operator conjugate to $\hat{\delta}$, and $n_g$ is a gate-charge offset. 
This 
Hamiltonian describes the degrees of freedom of the qubit. Further approximation steps lead to the
 models for different types of qubits. 
The ratio
$E_J/E_C$ determines the qubit regime: charge qubits operate at
$E_J/E_C \lesssim 1$, transmon qubits~\cite{koch2007} at $E_J/E_C \gg 1$, and flux
qubits in loops with multiple junctions. In every case, the nonlinearity
originates entirely from the Josephson junction.

Despite the experimental success of this circuit-based paradigm, current models
of superconducting quantum processors still face significant \emph{fidelity}
problems. Coherence times and gate fidelities have improved dramatically and now approach thresholds compatible with
surface-code error correction, yet they still lag behind the best trapped-ion
platforms in isolation~\cite{devoret2013,wendin2017,google2025,smith2025,ransford2026}.
Fidelity alone, however,
does not settle the comparison between hardware lines. Trapped-ion devices are
typically more rigid in their architecture, harder to scale and operate as
integrated processors, and exhibit single-qubit and two-qubit gate times, as well
as overall interaction timescales, that can be orders of magnitude longer than
those currently achieved with superconducting qubits. These operational
limitations compete directly with the coherence advantages of ions, so
that---when speed, integrability, and control overhead are taken
together---trapped-ion platforms do not presently enjoy a clear overall advantage
over superconducting circuits. Other commercial quantum technologies exist, but
the best-known large-scale processors developed by leading technology
companies---including Google's superconducting processors and IBM's integrated
quantum systems---rely on superconducting qubits rather than on ions. Improving
the fidelity of superconducting device models is therefore an important step
towards quantum computational advantage---the regime often referred to as quantum
supremacy. 

One possible explanation for the difference in operation fidelity between ion-trap-based
and superconducting processors is that the models of trapped ions are closer to the
underlying microscopic physics. By contrast, the standard route to superconducting-qubit
models outlined above involves three distinct steps: a microscopic many-body theory,
formulated in second quantisation, is used to derive the macroscopic behaviour of the
junction; this behaviour is then encoded in a classical lumped-element circuit; and the
circuit is finally quantised, yielding a quantum-mechanical description that is subject to
further approximations. The cumulative impact of these steps on the predicted dynamics is
difficult to assess systematically.
In practice, circuits with many elements are modelled by quantising the normal modes of
the linearised classical circuit, obtained from electromagnetic simulations, and treating
the junction nonlinearity perturbatively, as in black-box
quantisation~\cite{nigg2012} or the energy-participation
approach~\cite{minev2021,yilmaz2026}; see~\cite{ciani2024,rasmussen2021} for an
overview, and~\cite{osborne2024,parrarodriguez2024} for recent rigorous formulations of
circuit quantisation.
There are continuous and successful efforts to improve the fidelity of these effective
models, see e.g.~\cite{park2024accuracy}. Notably, these refined effective models are
already highlighting the need to correct the Josephson relations themselves: measured
transmon spectra are only reproduced accurately once higher harmonics are added to the
$\cos\delta$ potential of Eq.~\eqref{eq:qubit-hamiltonian}, that is, once the sinusoidal
current--phase relation~\eqref{eq:josephson-current} is
corrected~\cite{willsch2024,kim2026}. 
All these approaches rely, to a certain extent, on deriving a quantum Hamiltonian from a
classical circuit. Only recently have
approaches appeared that derive circuit Hamiltonians directly from a microscopic
description, bypassing the classical circuit~\cite{liao2025}.

This motivates
the central question addressed in the present work: \emph{to what extent can the
physics of a Josephson junction be captured by a purely quantum-mechanical model,
formulated directly in Hilbert space, without passing through an intermediate
classical circuit description?}

In this article we present a purely quantum-mechanical description of a simple circuit including a
Josephson junction that correctly recovers the macroscopic behaviour of the junction, 
Eqs.\ \eqref{eq:josephson-current} and \eqref{eq:josephson-voltage}. It provides an alternative
modelling tool for superconducting qubits that is constructed entirely from first
principles, and that we expect to improve operation fidelities.

Our main result concerns the simplest such circuit, a superconducting loop interrupted
by a single junction. We show that the ground-state momentum expectation, which we
identify with the supercurrent, factorises exactly as
\begin{equation*}
\langle\hat{p}_0\rangle = I_c(\delta,\lambda)\,\sin\delta,
\end{equation*}
where $\delta$ is the phase jump across the junction and $\lambda$ a coupling strength
that controls its transparency (Section~\ref{sec:first_josephson_relation}). For an opaque
junction, $\lambda\to-\infty$, the coefficient becomes independent of the phase,
$I_c\simeq\pi^2/(2|\lambda|)$, and the first Josephson relation~\eqref{eq:josephson-current}
is recovered. For strongly attractive coupling, $\lambda\gg1$, where the ground state is
bound to the junction, the $\sin\delta$ law is also recovered, with an exponentially
suppressed amplitude. Away from these limits, however, $I_c$ inherits a dependence on
$\delta$, which already appears at order $\lambda^{-2}$, and the model predicts
deviations from the pure sinusoidal law, cf.\ Eq.~\eqref{eq:ring-Ic-series}. For a fully transparent contact the current
becomes linear in $\delta$. These deviations are compatible with the higher Josephson
harmonics that the effective models discussed above incorporate phenomenologically in
order to reproduce measured spectra~\cite{willsch2024,kim2026}. Here, instead, they arise
as a \emph{prediction} of the model, and they grow with the transparency of the
junction, in qualitative agreement with the interpretation of~\cite{willsch2024} in terms
of high-transparency conduction channels. Finally, we show that when $\delta$ varies in
time the second Josephson relation~\eqref{eq:josephson-voltage} also emerges without
additional input (Section~\ref{sec:second_josephson_relation}).

\subsection{Mathematical Modelling of the Junction}
\label{sec:intro-modelling}

In this subsection we introduce the main assumption underlying the model.
A natural starting point is the \emph{point-contact limit}, in which the junction
width is much smaller than every other length scale in the problem. The junction
then reduces to a single point at which two one-dimensional conductors meet. The
dynamics on each side of the point is free (governed by the Laplacian), and the
entire effect of the junction is encoded in the matching conditions imposed on the
wave function and its derivative at the contact. These matching conditions must be
chosen so that the resulting Hamiltonian is self-adjoint, guaranteeing unitarity
of the time evolution and a real energy spectrum.

The mathematical framework for classifying all such boundary conditions is the
theory of \emph{self-adjoint extensions} of symmetric operators, introduced by von
Neumann~\cite{vonneumann1932} and developed in its general form for differential
operators by G.~Grubb~\cite{grubb1968}. In this
work we shall use the approach following Kochubei~\cite{kochubei1975}, Asorey,
Ibort, and Marmo~\cite{asorey2005}, and recent results using quadratic forms by
Ibort, Lled\'o, and
P\'erez-Pardo~\cite{ibortlledojfa2015,ibort2015,ibortlledo2015}. In this
framework, the space of self-adjoint extensions of the Laplacian on a split
interval is parametrised by the boundary data, and
every physically admissible junction condition corresponds to a specific choice of
the parameters.
Within this classification, the Josephson junction can be identified with a
particular two-parameter family of extensions, specified by a coupling strength
and a Josephson phase; the precise form of these boundary conditions is introduced
in Section~\ref{sec:jj-model}.

The rest of the article is organised as follows. Section~\ref{sec:jj-model} introduces the
one-particle model and the Josephson junction boundary conditions, discusses the meaning
of the parameters $\delta$ and $\lambda$, and shows that the phase jump is
gauge-equivalent to a magnetic flux threading the loop. Section~\ref{sec:ring} solves the
spectral problem on a ring with a single junction, in closed form for $\delta=0$ and for
$\lambda=0$ and numerically otherwise, including the bound states that appear for
$\lambda>0$. Section~\ref{sec:first_josephson_relation} computes the ground-state momentum
expectation, derives the factorisation $\langle\hat{p}_0\rangle=I_c\sin\delta$, and
analyses its asymptotic regimes and the deviations from the $\sin\delta$ law.
Section~\ref{sec:second_josephson_relation} considers a time-dependent phase and derives the
second Josephson relation. Section~\ref{sec:conclusions} summarises our results and
outlines future work.

\section{An Alternative Mathematical Description of the Josephson Junction}
\label{sec:jj-model}


\subsection{Introduction to the Model}
\label{sec:intro-model}

The time evolution of a closed quantum system is governed by the Schr\"odinger
equation. For this evolution to be \emph{unitary}---so that probability is
conserved---the Hamiltonian must be a self-adjoint operator on the Hilbert space
of states, as established by Stone's theorem, which identifies strongly continuous
one-parameter unitary groups with self-adjoint generators. For bounded operators,
the symmetry of the Hamiltonian ($\langle\Phi, H\Psi\rangle = \langle H\Phi, \Psi\rangle$) already
implies self-adjointness; for the unbounded differential operators relevant here,
however, symmetry is necessary but not sufficient. One must additionally specify a
\emph{domain} on which the operator acts, and different domains define physically
distinct dynamics. The classification of all self-adjoint extensions of a given
symmetric operator was characterised by von Neumann~\cite{vonneumann1932}. For the
Laplacian on a bounded domain, the familiar Dirichlet, Neumann, and Robin boundary
conditions each determine a self-adjoint realisation, but they are by no means the
only admissible possibilities; see~\cite{albeverio2005,bruning2008,asorey2005,ibortlledojfa2015} for general
treatments and~\cite{balmaseda2019,BalmasedaLonigroPerezPardo2023,angelone2023,cunden2026}
for further examples.

Superconducting materials realise macroscopic quantum coherence that can be
understood, at the many-body level, as a Bose--Einstein condensate of Cooper pairs
occupying a single macroscopic quantum state. A natural first approximation to the
full many-body problem is to study \emph{one-particle} Hamiltonians for the
condensate wave function, which is expected to be accurate in the
low-carrier-density regime. This is the approach adopted in this work. 
We will therefore postulate a one-particle Hamiltonian compatible with the physical structure
of the junction. The ground state of this one-particle Hamiltonian shall describe the physics
of the junction in the low-carrier-density limit, and the macroscopic observables shall be
related to the expectation values of the observables in this ground state.

In the remainder of this section we introduce a one-dimensional effective model of a Josephson
junction in which the contact is described by a point-like interface: the wave
function may undergo a phase jump and a discontinuity in its derivative at the
junction. The bulk dynamics on each side is governed by the (negative) Laplacian,
and the interface conditions are chosen so that the resulting operator is
self-adjoint.

\subsection{Boundary Conditions for the Josephson Junction}
\label{sec:BC}

The Laplacian $\Delta = -\frac{\mathrm{d}^2}{\mathrm{d} x^2}$ describes the 
free propagation of a quantum
particle on a one-dimensional region. In the context of superconductivity, it
provides a one-particle description of excitations in the condensate. A Josephson
junction introduces a discontinuity in this propagation at the point where the junction is
located, $x=0$, which can be
encoded entirely in the matching conditions at the contact. Following Balmaseda,
Di Cosmo, and P\'erez-Pardo~\cite{balmaseda2019}, we seek boundary conditions that
define a self-adjoint extension of the Laplacian and preserve the \emph{modulus} of
the wave function across the junction.

The simplest choice is \emph{continuity} at the junction:
$\psi(0^-) = \psi(0^+)$ and $\psi'(0^-) = \psi'(0^+)$, which corresponds to the
Laplacian with periodic boundary conditions at the junction, that is, to the absence of a
junction.
For a Josephson junction we relax this: we require continuity of the \emph{modulus} of
the wave function but allow a \emph{phase jump} $\delta \in \mathbb{R}$, and we
allow a \emph{jump in the derivative} controlled by a real parameter $\lambda$.
This is encoded in the following \emph{Josephson junction boundary conditions} at
$x = 0$:
\begin{subequations}
\label{eq:JJ-BC}
\begin{align}
\psi(0^-) &= e^{i\delta} \psi(0^+), \quad \delta \in \mathbb{R}, \label{eq:JJ-BC-a} \\
\psi'(0^-) - e^{i\delta} \psi'(0^+) &= \lambda \, \psi(0^-), \quad \lambda \in \mathbb{R}. \label{eq:JJ-BC-b}
\end{align}
\end{subequations}
Equations~\eqref{eq:JJ-BC-a} and~\eqref{eq:JJ-BC-b} are, according
to~\cite{balmaseda2019}, the most general boundary conditions that yield a
self-adjoint operator and preserve $|\psi|$ across the junction. The number of free
parameters in this model coincides with the number of experimentally
measurable quantities of the junction---the superconducting phase difference and
the critical current.
Equation~\eqref{eq:JJ-BC-a} implies $|\psi(0^-)| = |\psi(0^+)|$, so the probability
density is continuous across the junction, while the phase may jump by $\delta$.
Equation~\eqref{eq:JJ-BC-b} fixes the jump in the derivative in terms of
$\psi(0^-)$ and the real parameter $\lambda$. 
These boundary conditions define a self-adjoint operator, cf.~\cite{asorey2005,ibortlledojfa2015,ibort2015},
and therefore make the
problem well posed in the quantum-mechanical sense.

The two real parameters $\delta$ and $\lambda$ admit a direct physical
interpretation in terms of the Josephson junction they model. The phase $\delta$
is the macroscopic phase difference across the junction: it is the difference
between the phases of the order parameters (macroscopic wave functions) of the two
superconducting electrodes, and 
shall be identified with the variable $\delta$ that appears in the first Josephson 
relation $I = I_c\sin\delta$~\cite{josephson1962}. The parameter
$\lambda$ controls the strength of the derivative discontinuity and is related to
the tunnelling properties of the barrier. Indeed, multiplying $\psi$ by $e^{i\delta}$
on one side of the junction removes the phase from~\eqref{eq:JJ-BC} and leaves a
point interaction of strength $-\lambda$, 
with a transmission coefficient that decreases with $|\lambda|$
: the limit $\lambda = 0$ corresponds to a perfectly transparent contact (no
junction), while $|\lambda| \to \infty$ describes an opaque barrier that decouples
the two sides.

These boundary conditions admit generalisations to arbitrary circuits in which several 
Josephson junctions can be combined with additional external electric potentials
or magnetic fields. In all cases, assuming that the circuit does not contain 
wires of infinite length and that the number of junctions is finite, it can be proven 
that the spectrum is semibounded from below, not necessarily positive, and the eigenvectors
provide a complete orthonormal basis~\cite{BalmasedaLonigroPerezPardo2023}. In particular,
for $\lambda<0$ the operator, and therefore its spectrum, is positive, whereas for
$\lambda>0$ negative eigenvalues may appear.

\subsection{Relation of the model to external magnetic fields}\label{sec:planar_rotor}

There is a close relationship between the system so described and a system in which a
magnetic field threads the loop. We present this relation in the simple geometry
that we are describing, in which the parameter $\delta$ is directly related to the
magnetic vector potential.

To make this explicit, let us parametrise the loop by its arc length $x\in(-\ell,\ell)$,
so that its circumference is $2\ell$, and place the junction at the point where both
ends meet, $-\ell\sim\ell$, so that $\psi(0^-)=\psi(\ell)$ and $\psi(0^+)=\psi(-\ell)$
in~\eqref{eq:JJ-BC}.
Consider the unitary multiplication operator
\begin{equation}
(\mathcal{U}_\delta\psi)(x) = e^{-i\delta x/(2\ell)}\,\psi(x).
\label{eq:gauge-transformation}
\end{equation}
A direct computation shows that $\psi$ satisfies the Josephson
junction boundary conditions~\eqref{eq:JJ-BC} if and only if $\phi=\mathcal{U}_\delta\psi$ satisfies
\begin{equation}
\phi(\ell)=\phi(-\ell), \qquad \phi'(\ell)-\phi'(-\ell)=\lambda\,\phi(\ell),
\label{eq:periodic-delta-BC}
\end{equation}
that is, the phase jump disappears while the jump in the derivative is preserved.
Moreover,
\begin{equation}
\mathcal{U}_\delta\Bigl(-\frac{\mathrm{d}^2}{\mathrm{d}x^2}\Bigr)\mathcal{U}_\delta^{-1}
= \Bigl(-i\frac{\mathrm{d}}{\mathrm{d}x}+\frac{\delta}{2\ell}\Bigr)^{2},
\label{eq:magnetic-laplacian}
\end{equation}
which is the Hamiltonian of a charged particle on the loop subject to the constant
vector potential $A=-\delta/(2\ell)$, in units where $\hbar$ and the charge are equal to
one. Therefore, the Hamiltonian defined by the Josephson junction boundary
conditions~\eqref{eq:JJ-BC} is unitarily equivalent to the magnetic
Hamiltonian~\eqref{eq:magnetic-laplacian} with the boundary
conditions~\eqref{eq:periodic-delta-BC}, namely periodic boundary conditions together
with a point-like interaction of strength $\lambda$ located at the junction. The flux
of the vector potential through the loop is $\oint A\,\mathrm{d}x=-\delta$; restoring
units, $|\delta| = 2\pi|\Phi_B|/\Phi_0$, where $\Phi_B$ is the magnetic flux threading
the loop and $\Phi_0=h/q$ is the flux quantum for carriers of charge $q$, which for
Cooper pairs ($q=2e$) is the superconducting flux quantum $h/2e$.

For $\lambda=0$, conditions~\eqref{eq:periodic-delta-BC} are purely periodic and,
after rescaling the loop to unit radius, \eqref{eq:magnetic-laplacian} becomes the
Hamiltonian of the planar rotor threaded by a magnetic flux studied by Asorey, Esteve,
and Pacheco~\cite[Sec.~II.A]{asorey1983}. There, the same gauge transformation relates
the magnetic description with periodic boundary conditions to the free rotor with
quasi-periodic boundary conditions, the phase being proportional to the flux; both
descriptions share a spectrum that differs from that of the free rotor unless the flux
is an integer multiple of the flux quantum. The case $\lambda\neq 0$, as well as its
generalisation to networks of higher-dimensional manifolds, follows from the results
of~\cite{BalmasedaLonigroPerezPardo2023}: magnetic Laplacians with an exact vector
potential are unitarily equivalent, through a gauge transformation of the
form~\eqref{eq:gauge-transformation}, to Laplacians with correspondingly transformed
boundary conditions, and conditions of the type~\eqref{eq:JJ-BC} belong to the class of
quasi-$\delta$ boundary conditions considered there.

\subsection{Momentum expectation}
\label{sec:momentum-expectation}

Given a state of the one-dimensional quantum system, the macroscopic observable relevant
to the first Josephson relation is the electric current, which we identify with
the expectation value of the momentum operator,

$$I = \frac{q}{m}\langle \hat{p} \rangle.$$

For a charged particle moving on the one-dimensional loop with the Josephson junction
boundary conditions~\eqref{eq:JJ-BC}, the momentum operator
$\hat{p}=-i\hbar\,\frac{\mathrm{d}}{\mathrm{d}x}$ is essentially
self-adjoint, and the above expectation value is well defined\footnote{Notice that
this is not the case for an open loop, i.e.\ an interval, where the momentum operator does not have self-adjoint
realisations.}.


\section{Spectrum of a Josephson Junction on a Ring}
\label{sec:ring}


We now solve the spectral problem for a single junction placed on a \emph{closed
ring} of circumference~$2$, the geometry relevant to superconducting loops. To simplify
the exposition we will consider natural units and, since $\ell$ only sets the length
scale, we fix $\ell=1$ in the parametrisation of Section~\ref{sec:planar_rotor}.
Topologically, the ring is the interval $[-1,1]$ with its endpoints $-1$ and $1$
\emph{identified}, $-1\sim 1$. We place the junction precisely at the
identification point, so that the configuration space is
$\Omega^{\mathrm{ring}} = (-1,1)$ and all the matching data of the contact are
carried by the boundary equations at $x = \pm 1$. 

The free Hamiltonian is
$-\mathrm{d}^2/\mathrm{d}x^2$ acting (initially) on the Sobolev space of order~2, $H^2(-1,1)$,
subject to the following boundary conditions.
The two limits
$\Psi(1^-)$ and $\Psi(-1^+)$, the values of the wave function on the two sides of
the identified point, now play the role of $\psi(0^-)$ and $\psi(0^+)$ of
Section~\ref{sec:jj-model}. 
We therefore impose the boundary conditions
\begin{subequations}
\label{eq:ring-junction-pm1}
\begin{align}
\Psi(1) &= e^{i\delta}\,\Psi(-1), \label{eq:ring-junction-pm1-a} \\
\Psi'(1) - e^{i\delta}\,\Psi'(-1) &= \lambda\,\Psi(1), \label{eq:ring-junction-pm1-b}
\end{align}
\end{subequations}
which are the verbatim translation of~\eqref{eq:JJ-BC} under the relabelling
$\psi(0^-) \leftrightarrow \Psi(1)$, $\psi(0^+) \leftrightarrow \Psi(-1)$. 

In particular, taking
$\delta = \lambda = 0$ in~\eqref{eq:ring-junction-pm1} reproduces the standard
\emph{periodic} (Kirchhoff--Neumann) conditions $\Psi(1) = \Psi(-1)$,
$\Psi'(1) = \Psi'(-1)$ of a free particle on a ring~\cite{berkolaiko2013}, so the
present setup is the natural two-parameter family $(\delta,\lambda)$ of
self-adjoint extensions interpolating between the free ring and an opaque
(large-$|\lambda|$) point-like interaction where a phase discontinuity is allowed.
In the case $\delta\neq 0$ and $\lambda = 0$ we recover the situation of the 
planar rotor described at the end of Section~\ref{sec:planar_rotor}.

\subsection{Secular equation}
\label{sec:ring-2x2}

The eigenvalue equation $-\Psi'' = E\Psi$ has the general plane-wave solution
\begin{equation}
\Psi(x) = A\,e^{ikx} + B\,e^{-ikx}, \qquad x\in(-1,1),
\label{eq:ring-ansatz-2x2}
\end{equation}
for $E=k^2$,
parametrised by two complex coefficients $(A,B)$. 
The self-adjointness of the Hamiltonian guarantees that the spectrum is real but not
necessarily positive. Hence $E=k^2\in\mathbb{R}$ implies that $k$ is either real
or purely imaginary, corresponding, respectively, to positive and negative
eigenvalues.
Plugging~\eqref{eq:ring-ansatz-2x2}
into the two Josephson conditions~\eqref{eq:ring-junction-pm1} gives a homogeneous
linear system $M^{\mathrm{ring}}(k;\delta,\lambda)\,\mathbf{w} = \mathbf{0}$ on
$\mathbf{w} = (A,B)^{\mathsf{T}}$. Introducing the shorthand
\begin{equation}
\alpha := e^{ik} - e^{i\delta}e^{-ik}, \qquad \beta := e^{-ik} - e^{i\delta}e^{ik},
\label{eq:ring-alphabeta}
\end{equation}
the matrix takes the compact form
\begin{equation}
M^{\mathrm{ring}} =
\begin{pmatrix}
\alpha & \beta \\[3pt]
ik\,\alpha - \lambda e^{ik} & -ik\,\beta - \lambda e^{-ik}
\end{pmatrix}.
\label{eq:M-ring-2x2}
\end{equation}
Expanding the determinant gives
$\det M^{\mathrm{ring}} = -2ik\,\alpha\beta + \lambda(\beta\,e^{ik} - \alpha\,e^{-ik})$.
The trigonometric identities
$\alpha\beta = 2\,e^{i\delta}\,(\cos\delta - \cos 2k)$ and
$\beta\,e^{ik} - \alpha\,e^{-ik} = -2i\,e^{i\delta}\sin 2k$, collapse this to
\begin{equation}
\det M^{\mathrm{ring}} = -2i\,e^{i\delta}\,\bigl[\,2k(\cos\delta - \cos 2k) + \lambda\,\sin 2k\,\bigr].
\label{eq:ring-det-2x2-closed}
\end{equation}
The prefactor $-2i\,e^{i\delta}$ never vanishes, so $\det M^{\mathrm{ring}}=0$ is
equivalent to the transcendental equation
\begin{equation}
F(k;\delta,\lambda) := 2k\bigl(\cos\delta - \cos 2k\bigr) + \lambda\,\sin 2k = 0.
\label{eq:ring-scalar-secular}
\end{equation}
Notice that this equation is antisymmetric under the change $k\to-k$ and the corresponding 
solutions, Eq.\ \eqref{eq:ring-ansatz-2x2}, are linearly dependent. Hence it is enough to explore
the positive axis and the positive imaginary axis. Also notice that $k=0$ corresponds
to the constant function, which is a solution only in the trivial case $\lambda=\delta=0$.

\subsection{Spectrum}
\label{sec:ring-spectrum}

The secular equation~\eqref{eq:ring-scalar-secular}
admits closed-form solutions along the the axis $\lambda=0$ (the planar rotor); in the
generic interior of the $(\delta,\lambda)$ plane the roots are
obtained numerically. We first explore the positive part of the spectrum.

\subsubsection{The case \texorpdfstring{$\delta=0$}{delta=0}}
\label{subsec:ring-delta0}

For $\delta=0$, using $1-\cos 2k = 2\sin^{2}k$ and $\sin 2k = 2\sin k\cos k$,
Eq.~\eqref{eq:ring-scalar-secular} factorises as
$F(k;0,\lambda) = 2\sin k\,[\,2k\sin k + \lambda\cos k\,]$, yielding two independent
root families. The \emph{bare ring modes} from $\sin k = 0$,
\begin{equation}
k_n^{(\mathrm{p})} = n\pi, \qquad E_n^{(\mathrm{p})} = n^{2}\pi^{2},
\qquad n = 0,1,2,\ldots,
\label{eq:ring-naked}
\end{equation}
are \emph{independent of} $\lambda$: standing waves with a node at $x=\pm1$, on which
the junction strength is invisible. The case $n=0$ provides a non-trivial solution only
if $\lambda = 0$. 
The \emph{junction-dressed modes}
from $2k\sin k + \lambda\cos k = 0$, i.e.\
\begin{equation}
\lambda = -2k\tan k,
\label{eq:ring-tan}
\end{equation}
contribute exactly one $\lambda$-dependent root per branch of the tangent function. At $\lambda=0$ the two
families coincide on $k=n\pi$ (a double root of $F$), recovering the free-ring
spectrum~\eqref{eq:ring-naked}; for $\lambda\neq 0$ the double roots split into two
simple roots and the level density doubles, except for the ground state, $n=0$, which
remains simple because the bare ring modes do not provide a non-trivial solution for $n=0$.

\subsubsection{Splitting of the ring levels (\texorpdfstring{$\delta\neq 0$}{delta !=0} at \texorpdfstring{$\lambda=0$}{lambda=0})}
\label{subsec:ring-AB}

For arbitrary $\delta$ at $\lambda=0$, $F$ reduces to $2k(\cos\delta - \cos 2k)$.
Using $\cos\delta - \cos 2k = 2\sin(k+\delta/2)\sin(k-\delta/2)$, the positive roots
are
\begin{equation}
k_n^{\pm}(\delta) = n\pi \pm \tfrac{\delta}{2},
\qquad
E_n^{\pm}(\delta) = \bigl(n\pi \pm \tfrac{\delta}{2}\bigr)^{2},
\label{eq:ring-AB-shift}
\end{equation}
$n = 1, 2, \ldots$: the doubly degenerate periodic levels
$k_n^{(\mathrm{p})} = n\pi$ split linearly in $\delta$ into two interleaved ladders
symmetric about $n\pi$. A non-zero Josephson phase therefore lifts the degeneracy of
each bare mode on the ring, recovering the spectrum of the planar rotor~\cite{asorey1983}.

\subsubsection{Bound states (\texorpdfstring{$E<0$}{E<0})}
\label{subsec:ring-bound}

Negative energies arise from the positive imaginary axis $k = i\kappa$ with $\kappa>0$,
turning~\eqref{eq:ring-scalar-secular} into
\begin{equation}
G(\kappa;\delta,\lambda) := 2\kappa\bigl(\cos\delta - \cosh 2\kappa\bigr) + \lambda\,\sinh 2\kappa = 0.
\label{eq:ring-bound-secular}
\end{equation}
Setting $\delta=0$ factorises
$$G(\kappa;0,\lambda) = 2\sinh\kappa\,[-2\kappa\sinh\kappa + \lambda\cosh\kappa],$$ 
so either $\kappa=0$ or $\lambda = 2\kappa\tanh\kappa$. Then for any $\lambda>0$ 
there is a negative-energy solution, as anticipated. 
Conversely, at $\lambda=0$ one has $\cosh 2\kappa>1\ge\cos\delta$
for $\kappa>0$, so \emph{without junction coupling no hyperbolic bound state sits at
$E<0$}. For the generic case $\kappa>0$, Eq.~\eqref{eq:ring-bound-secular} is
equivalent to $\lambda\sinh 2\kappa = 2\kappa(\cosh 2\kappa-\cos\delta)$. Both sides
vanish at $\kappa=0$ and increase monotonically, the right-hand side growing faster as
$\kappa\to\infty$. Hence, if at $\kappa=0$ the slope of the left-hand side, $2\lambda$,
exceeds that of the right-hand side, $2(1-\cos\delta)$, the two curves must cross at some
$\kappa>0$. This happens whenever
\begin{equation}
\lambda > \lambda_{\mathrm{th}}(\delta) := 1 - \cos\delta = 2\sin^{2}(\delta/2).
\label{eq:ring-bound-threshold}
\end{equation}
A non-zero Josephson phase thus stiffens the ring, requiring $\lambda$ to exceed
$2\sin^{2}(\delta/2)$ before a negative-energy state appears. This state is the analytic
continuation through $E=0$ of the lowest oscillatory mode, not a new degree of
freedom.

\subsubsection{Generic spectrum}
\label{subsec:ring-numerics}

\begin{figure}[t]
\centering
\includegraphics[width=\linewidth]{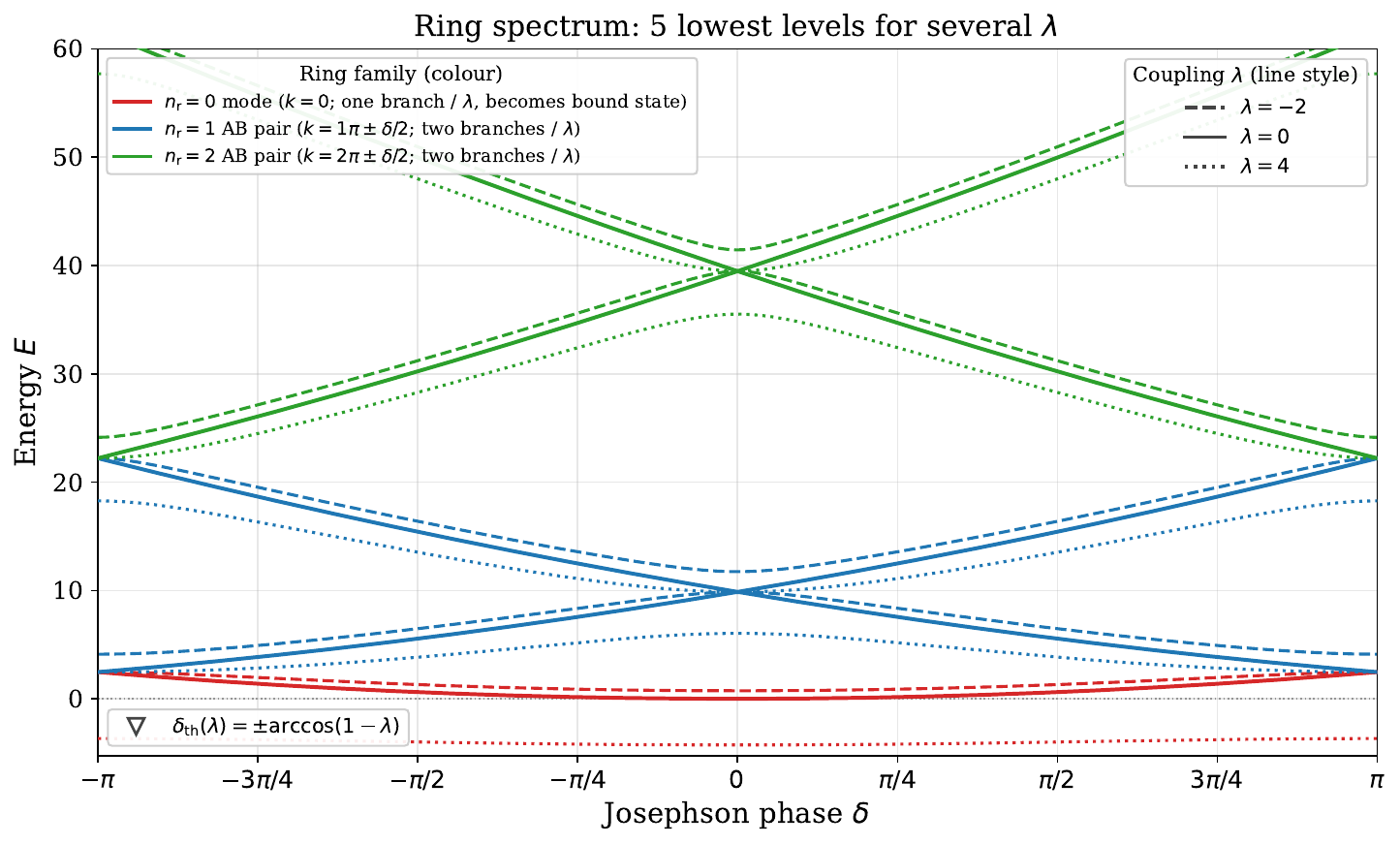}
\caption{Five lowest signed energies of the ring as functions of the Josephson
phase $\delta\in(-\pi,\pi)$, for $\lambda\in\{-2,0,4\}$ (line styles: dashed, solid,
dotted). Curves are coloured by ring-family index $n_{\mathrm{r}}$ (red:
$n_{\mathrm{r}}=0$; blue: $n_{\mathrm{r}}=1$; green: $n_{\mathrm{r}}=2$). Each
$n_{\mathrm{r}}\geq 1$ family consists of two branches
$k=n_{\mathrm{r}}\pi\pm\delta/2$~\eqref{eq:ring-AB-shift}. For $\lambda=4>2$ the
$n_{\mathrm{r}}=0$ mode (red) is bound ($E_0<0$) for all $\delta$.}
\label{fig:ring-spectrum-vs-delta}
\end{figure}

\begin{figure}[t]
\centering
\includegraphics[width=\linewidth]{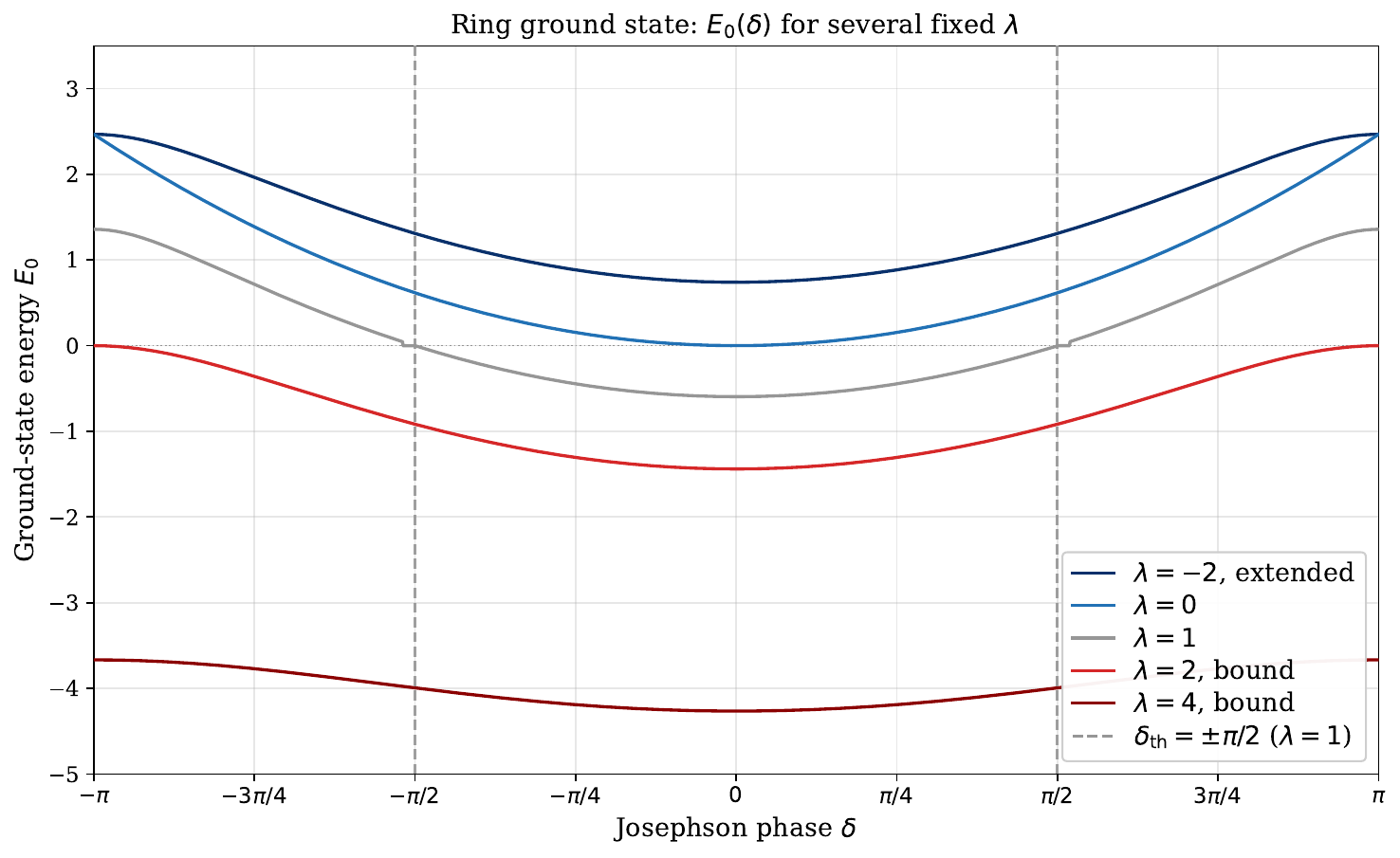}
\caption{Ground-state energy $E_0(\delta)$ for $\lambda\in\{-2,0,1,2,4\}$. For
$\lambda=1$ (grey), vertical dashed lines mark the bound-state threshold
$\delta_{\mathrm{th}}=\pm\pi/2$~\eqref{eq:ring-bound-threshold}. For $\lambda=2$
(red) the state is bound ($E_0\leq 0$) for all $\delta$, touching $E=0$ only at
$\delta=\pm\pi$; for $\lambda=4$ (dark red) it is deeply bound everywhere.}
\label{fig:ring-E0-vs-delta}
\end{figure}

For
prescribed $(\delta,\lambda)$ we will solve for $k$ in Eq.~\eqref{eq:ring-scalar-secular} numerically.
Roots of~\eqref{eq:ring-scalar-secular}
and~\eqref{eq:ring-bound-secular} are found by bracketing $F$ or $G$ on a fine grid
and refining by bisection. We track the signed energies as functions of the
Josephson phase $\delta$, the physically controlled parameter in a superconducting
loop experiment. Figure~\ref{fig:ring-spectrum-vs-delta} shows the five lowest
energies versus $\delta$ for $\lambda\in\{-2,0,4\}$.
They correspond to the reference values $n=0,1,2$ and we will label them, respectively, by an 
index $n_r=0,1,2$. 
Each $n_{\mathrm{r}}\geq 1$
family splits into two branches $k = n_{\mathrm{r}}\pi \pm \delta/2$
\eqref{eq:ring-AB-shift} that open symmetrically as $|\delta|$ grows, while the
$n_{\mathrm{r}}=0$ mode either stays positive ($\lambda=-2$) or dips below $E=0$
once $|\delta|$ crosses $\delta_{\mathrm{th}}(\lambda) = \arccos(1-\lambda)$, c.f.\ \eqref{eq:ring-bound-threshold}.
Figure~\ref{fig:ring-E0-vs-delta} isolates the ground-state energy $E_0(\delta)$
for $\lambda\in\{-2,0,1,2,4\}$: binding at fixed $\lambda$ occurs precisely when
$\lambda > \lambda_{\mathrm{th}}(\delta) = 1-\cos\delta$~\eqref{eq:ring-bound-threshold}.

\subsection{Eigenfunctions of the ring}
\label{sec:ring-eigenfunctions}

\begin{figure}[t]
\centering
\includegraphics[width=\linewidth]{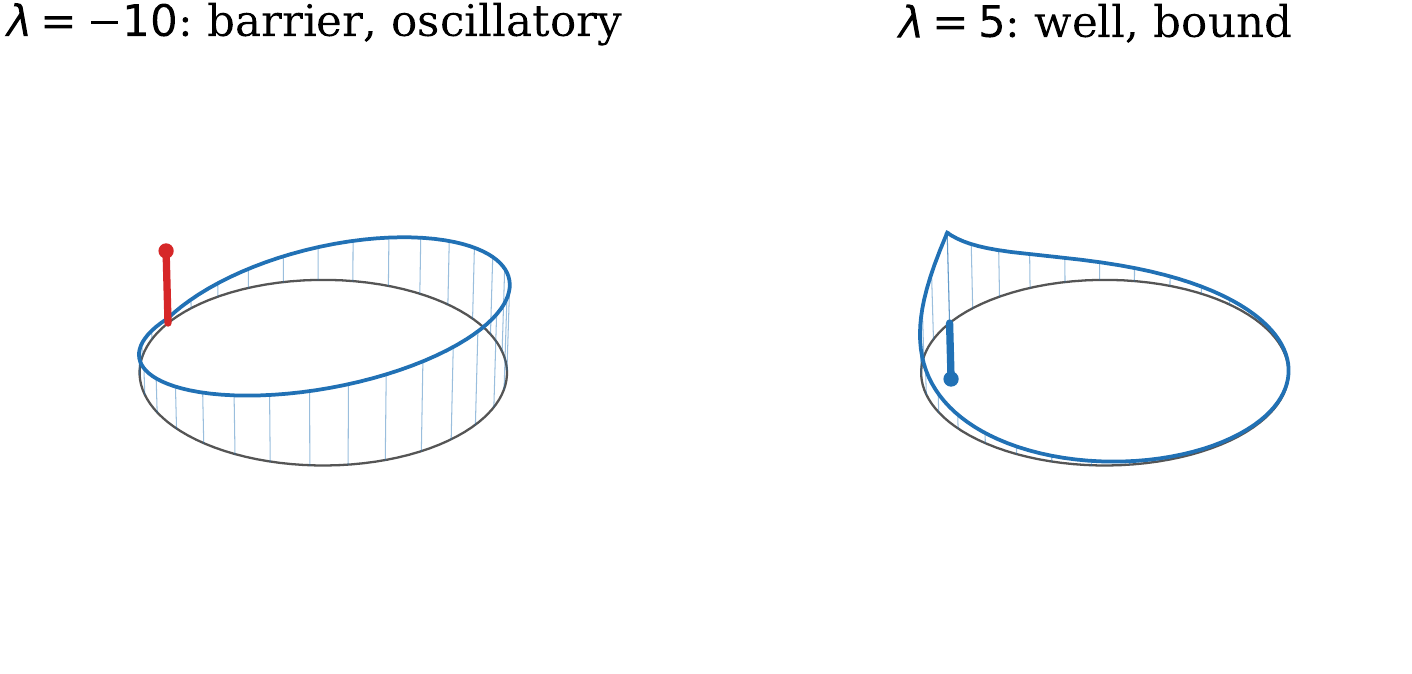}
\caption{Ground-state probability density $|\Psi_0(x)|^2$ as a height profile on the
ring, at $\delta=1$\,rad, for the two extreme regimes. \emph{Left} ($\lambda=-10$):
the junction acts as a \emph{barrier}, the density is depleted at the contact and
spreads over the bulk (oscillatory ground state, $E_0>0$). \emph{Right}
($\lambda=5$): the junction acts as a \emph{well}, the density concentrates at the
contact (bound ground state, $E_0<0$).}
\label{fig:ring-3d}
\end{figure}

Once~\eqref{eq:ring-scalar-secular} has been solved for a wave number $k_n>0$, the
corresponding 
eigenfunction follows from the kernel of $M^{\mathrm{ring}}$~\eqref{eq:M-ring-2x2}.
The first row reads $\alpha A + \beta B = 0$; using
$\alpha = 2 i\, e^{i\delta/2}\sin(k - \delta/2)$ and
$\beta = -2 i\, e^{i\delta/2}\sin(k + \delta/2)$, the amplitude ratio is
$B/A = \sin(k-\delta/2)/\sin(k+\delta/2)$. After normalising in $L^2(-1,1)$,
\begin{equation}
\Psi_n(x) = N_n\,\Bigl[\,\sin\!\bigl(k_n+\tfrac{\delta}{2}\bigr)\,e^{i k_n x} + \sin\!\bigl(k_n-\tfrac{\delta}{2}\bigr)\,e^{-i k_n x}\,\Bigr],
\label{eq:ring-eigenfunction}
\end{equation}
where $k_n$ is any positive root of $F(k;\delta,\lambda) = 0$. The addition formula
brings~\eqref{eq:ring-eigenfunction} to the manifestly real-and-imaginary form
\begin{equation}
\begin{split}
\Psi_n(x) = 2 N_n\,\Bigl[\,&\sin k_n\cos\!\tfrac{\delta}{2}\,\cos(k_n x)\\
&+ i\cos k_n\sin\!\tfrac{\delta}{2}\,\sin(k_n x)\,\Bigr].
\end{split}
\label{eq:ring-eigenfunction-realim}
\end{equation}
Setting $\delta=0$ kills the imaginary part, leaving a real cosine
$\Psi_n\propto\cos(k_n x)$. At $\lambda=0$ the amplitudes become equal in magnitude
and opposite in phase, reducing~\eqref{eq:ring-eigenfunction} to a single travelling
wave $e^{\pm ikx}$, as expected, recovering the 
eigenfunctions of the planar rotor. For the bound ground state ($E_0<0$, i.e.\
$\lambda>\lambda_{\mathrm{th}}(\delta)$), the substitution $k=i\kappa$
in~\eqref{eq:ring-eigenfunction-realim} gives
\begin{equation}
\begin{split}
\Psi_0(x) = N_0\,\Bigl[\,&-\sin\!\tfrac{\delta}{2}\cosh\kappa_0\,\sinh(\kappa_0 x)\\
&+ i\cos\!\tfrac{\delta}{2}\sinh\kappa_0\,\cosh(\kappa_0 x)\,\Bigr],
\end{split}
\label{eq:ring-bound-eigenfunction}
\end{equation}
which at $\delta=0$ reduces to a purely imaginary $\cosh(\kappa_0 x)$, peaked at the
junction. Therefore, when $\lambda>0$ the contact acts as a \emph{potential well} and the
density concentrates near $x=\pm 1$, where the junction is located.
When $\lambda<0$ it acts as a \emph{potential
barrier} and the density is depleted at the junction; at $\lambda=0$ the density is
uniform (Fig.~\ref{fig:ring-3d}).

\section{Ground-state momentum expectation and the Josephson relation}
\label{sec:first_josephson_relation}

\begin{figure}[t]
\centering
\includegraphics[width=\linewidth]{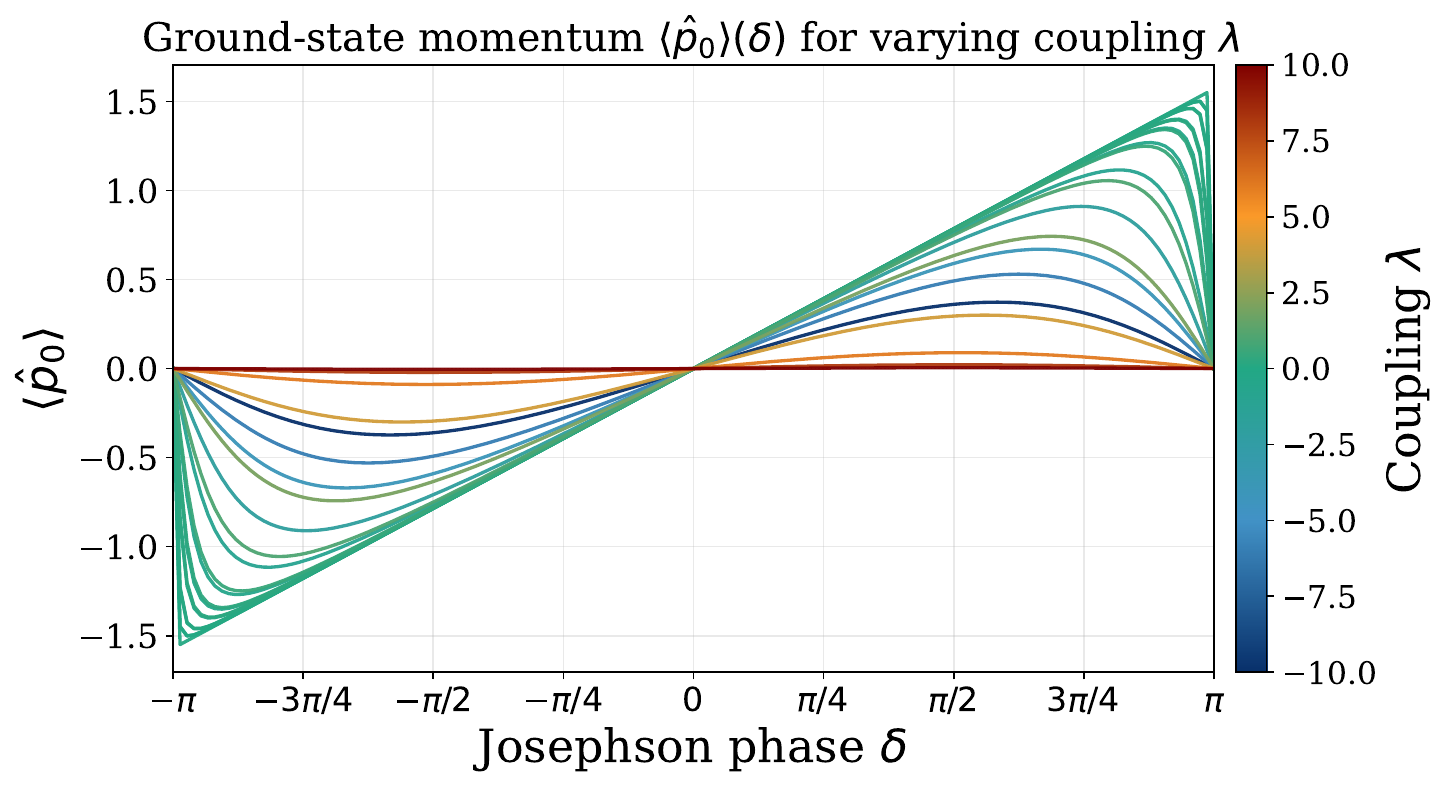}
\caption{Ground-state momentum expectation $\langle\hat{p}_0\rangle$ versus the
Josephson phase $\delta$, for a range of coupling strengths $\lambda$ (colour bar).
Blue: large negative $\lambda$ (oscillatory ground state, strong barrier); red:
large positive $\lambda$ (bound or near-bound ground state). For $|\lambda|\gg 1$
the curve is an almost pure $\sin\delta$; near $\lambda=0$ higher harmonics build a
sawtooth.}
\label{fig:p0-vs-lambda-sweep}
\end{figure}

Section~\ref{sec:momentum-expectation} introduced the expectation value of the 
momentum, 
$\langle\hat{p}\rangle$, as the key quantity related to 
the electrical current on the ring. On the ring the
ground state can carry $\langle\hat{p}_0\rangle\neq 0$. Our goal is to compute it as
a function of $(\delta,\lambda)$ and to recover the Josephson relation
$\langle\hat{p}_0\rangle=I_c\sin\delta$, in analogy with
Eq.~\eqref{eq:josephson-current}. Figure~\ref{fig:p0-vs-lambda-sweep} gives the
numerical overview: each curve shows $\langle\hat{p}_0\rangle(\delta)$ for one
coupling strength $\lambda$.

\subsection{Momentum expectation and the Josephson relation}
\label{subsec:ring-josephson-relation}

For $\lambda<\lambda_{\mathrm{th}}(\delta)$ (oscillatory ground state), the lowest
root of~\eqref{eq:ring-scalar-secular} is $k_0>0$. Using the
decomposition~\eqref{eq:ring-eigenfunction-realim}, introduce
\begin{equation}
a := \sin k_0\cos\!\tfrac{\delta}{2}, \qquad b := \cos k_0\sin\!\tfrac{\delta}{2},
\label{eq:ring-gs-amplitudes}
\end{equation}
so that $\Psi_0(x) = 2N_0[a\cos(k_0 x)+ib\sin(k_0 x)]$. Acting with
$\hat{p}=-i\partial_x$, the term odd in $x$ integrates to zero over $[-1,1]$ and only
the constant $ab$ contributes, giving
$\langle\hat{p}_0\rangle = (2N_0)^2\,2k_0\, ab$. The key identity is the
factorisation
\begin{equation}
ab = \tfrac{1}{4}\sin(2k_0)\,\sin\delta,
\label{eq:ring-ab-sind}
\end{equation}
which displays an \emph{explicit} factor of $\sin\delta$. The ground-state wave
number $k_0$ is itself a root of $F(k;\delta,\lambda)=0$ and therefore depends
implicitly on $\delta$. With the normalisation
$$(2N_0)^2 = 1/[a^2+b^2+(a^2-b^2)\sin(2k_0)/(2k_0)]$$
 we obtain
\begin{equation}
\langle\hat{p}_0\rangle = I_c(k_0;\delta,\lambda)\,\sin\delta,
\label{eq:ring-josephson-factor}
\end{equation}
with the Josephson coefficient
\begin{equation}
I_c(k_0;\delta,\lambda) := \frac{\tfrac{1}{2}k_0\sin(2k_0)}
{a^2+b^2+\dfrac{(a^2-b^2)\sin(2k_0)}{2k_0}}.
\label{eq:ring-Ic-formula}
\end{equation}
Equation~\eqref{eq:ring-josephson-factor} coincides with the first Josephson
relation~\eqref{eq:josephson-current} whenever $I_c$ is independent of $\delta$;
Figure~\ref{fig:p0-vs-lambda-sweep} shows that for $|\lambda|\gg 1$ the curves are
very close to a pure $\sin\delta$ law, while as $|\lambda|\to 0$ higher harmonics
build up into a sawtooth.

\paragraph{Free-ring limit ($\lambda\to 0$).}
At $\lambda=0$ the lowest positive root is $k_0=\delta/2$, corresponding to the free
plane wave $\Psi_0(x)\propto e^{i(\delta/2)x}$, for which
$\hat{p}_0\Psi_0 = (\delta/2)\Psi_0$ and hence
\begin{equation}
\langle\hat{p}_0\rangle\big|_{\lambda=0} = \frac{\delta}{2}.
\label{eq:ring-p0-free}
\end{equation}
The expectation is \emph{linear} in $\delta$ at zero coupling: it is the kinematic
momentum acquired by a particle moving on the ring in the presence of the phase
$\delta$, with no contribution from the (transparent) junction. In the unitarily
equivalent picture of the planar rotor, it is the current associated to the magnetic field
whose flux is proportional to $\delta$.

\subsection{Strong-coupling limit for a repulsive barrier}
\label{subsec:ring-josephson-strong}

\begin{figure}[t]
\centering
\includegraphics[width=\linewidth]{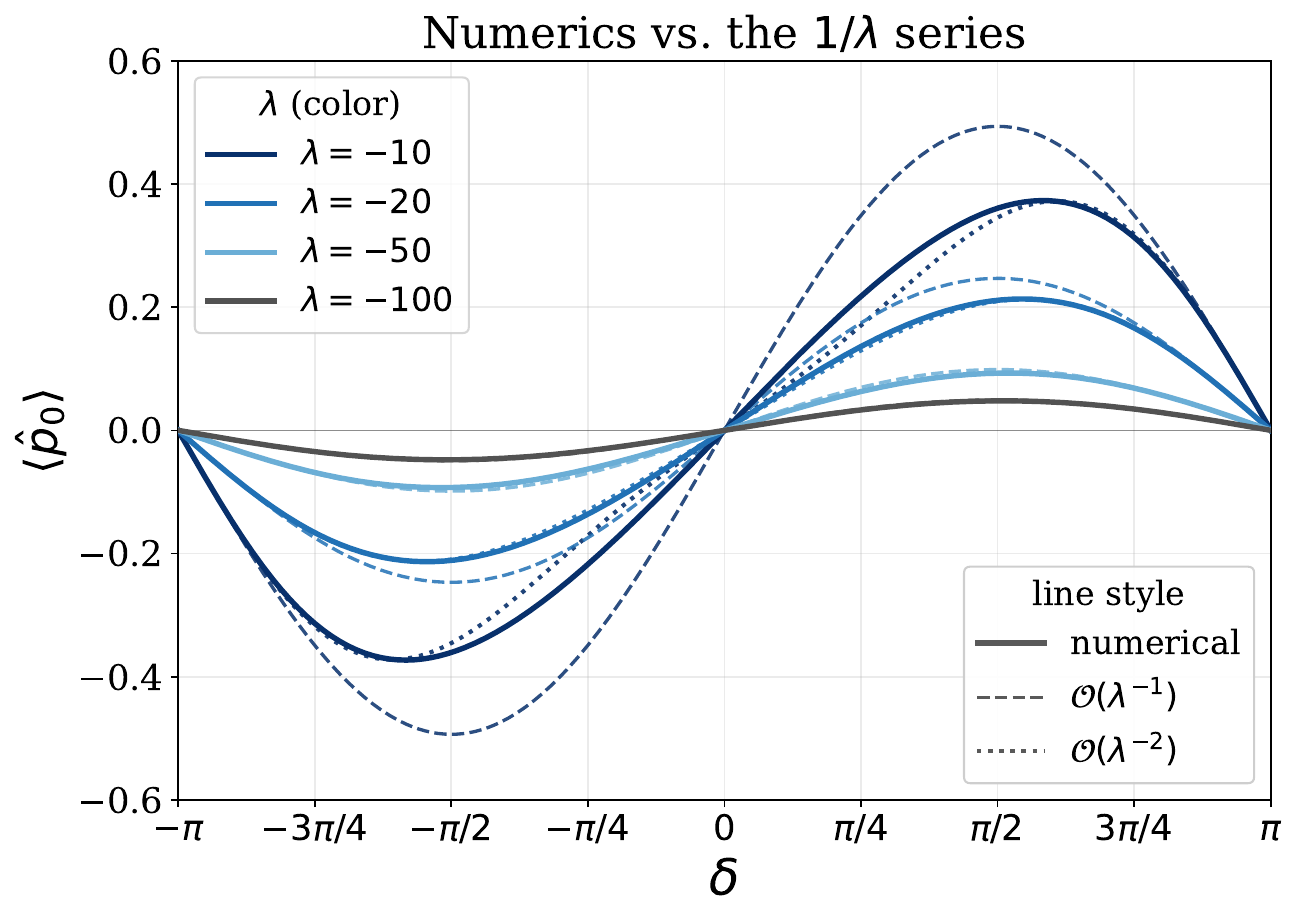}
\caption{Numerically computed $\langle\hat{p}_0\rangle$ (solid), obtained by
root-finding on~\eqref{eq:ring-scalar-secular}, vs.\ the large-coupling truncations
$\mathcal{O}(\lambda^{-1})$ (dashed) and $\mathcal{O}(\lambda^{-2})$ (dotted) of the
series~\eqref{eq:ring-Ic-series}, at $\lambda\in\{-10,-20,-50,-100\}$.}
\label{fig:p0-low-lambda}
\end{figure}

The factorisation~\eqref{eq:ring-josephson-factor} is exact but implicit, because
$k_0$ depends on $(\delta,\lambda)$. We now approximate this
root in the opaque-junction regime of strong coupling ($\lambda\to-\infty$), i.e.\
weak transmission.
In this regime the root satisfies 
$$2k(\cos\delta-\cos 2k) = |\lambda|\sin 2k.$$
Using Bolzano's theorem and the properties of the trigonometric functions, when $\delta\neq0$
there is one root in the interval $k\in(0,\pi/2)$, corresponding to the ground state.
Since $|\cos\delta-\cos 2k|\leq 2$ the root satisfies 
$$\frac{|\sin 2k|}{2k} \leq \frac{2}{|\lambda|}$$
and therefore, in the limit $\lambda\to-\infty$, the root must be near the first root 
of the $\operatorname{sinc} 2k$ function and the ground-state wave number $k_0$ approaches
$\pi/2$ from below. 
Writing $k=\tfrac{\pi}{2}+\varepsilon$ and expanding
$F$ to second order gives a quadratic in $\varepsilon$ whose root will approximate
the root of the secular equation. Only one of its roots is admissible and,
expanding its closed-form expression in powers of $1/\lambda$, one gets
$$\varepsilon = \pi(\cos\delta+1)/(2\lambda) + \pi(\cos\delta+1)^2/(2\lambda^2) + O(\lambda^{-3}).$$ 
Substituting
$k_0=\tfrac{\pi}{2}+\varepsilon$ into the implicit expression for the 
current~\eqref{eq:ring-Ic-formula}
and expanding up to order $O(\lambda^{-3})$ we get
\begin{equation}
I_c(k_0;\delta,\lambda) = \frac{J_0}{\lambda}+\frac{J_1(\delta)}{\lambda^{2}}+O(\lambda^{-3}),
\label{eq:ring-Ic-series}
\end{equation}
with $J_0=-\pi^2/2$ and $J_1(\delta)=-\frac{3}{2}\pi^2(\cos\delta+1)$, so the leading term is
$\delta$-independent and the $1/\lambda^2$ correction includes a phase
dependence. The overall suppression $I_c\sim|\lambda|^{-1}$
reflects the growing opacity of the junction. Figure~\ref{fig:p0-low-lambda} tests
this expansion: the two-term series is accurate for $|\lambda|\gtrsim 20$.

Therefore, the model recovers the Josephson current relation~\eqref{eq:josephson-current}
in the strong-coupling limit for a repulsive barrier, but also \emph{predicts a deviation}
from it at weaker coupling. 

\subsection{Strong-coupling limit for an attractive junction}
\label{subsec:ring-josephson-bound}

\begin{figure}[t]
\centering
\includegraphics[width=\linewidth]{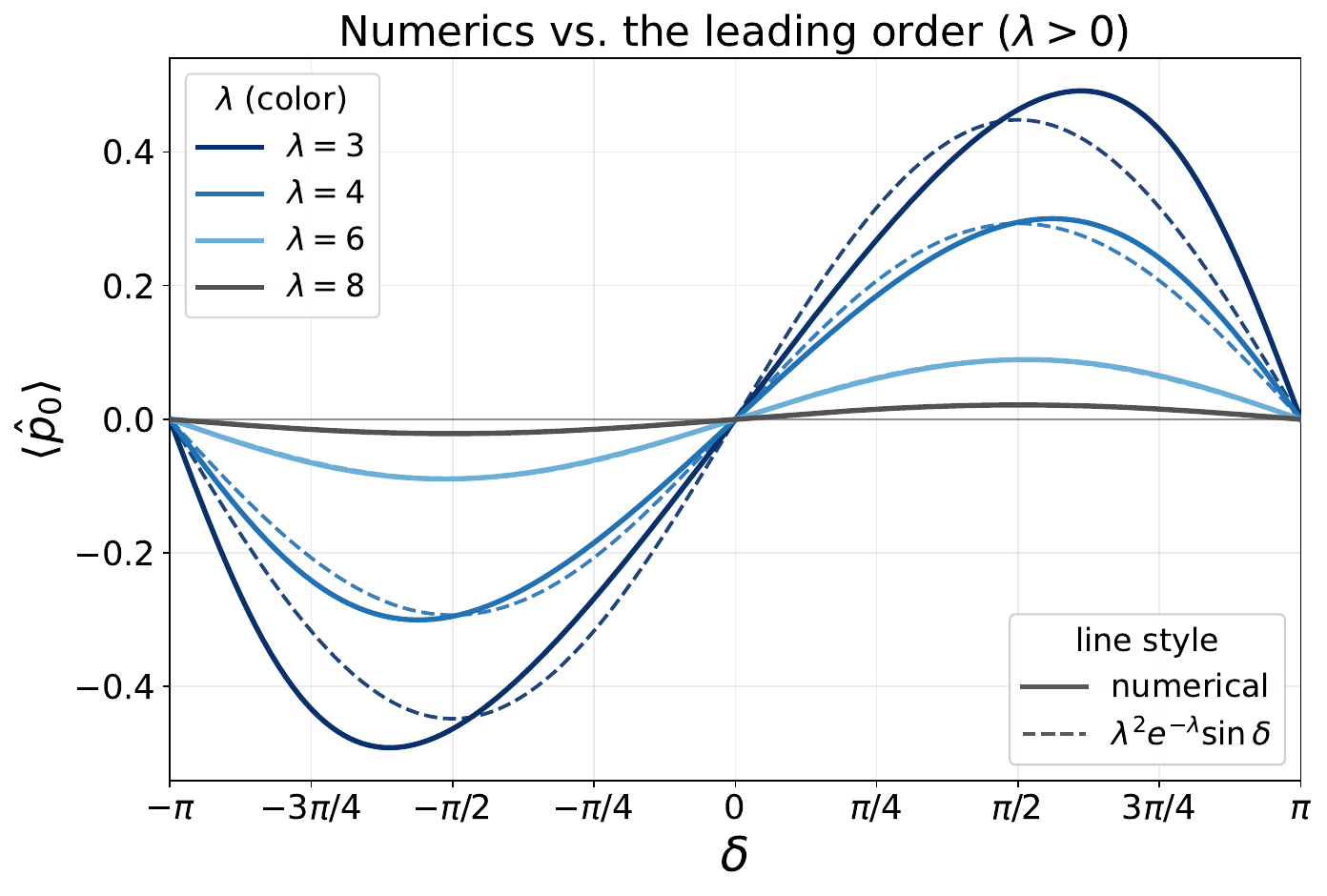}
\caption{Numerically computed $\langle\hat{p}_0\rangle$ (solid) vs.\ the large-coupling truncation
$\mathcal{O}(\lambda^2e^{-\lambda})$ (dashed)~\eqref{eq:ring-Ic-bound-limit}, at $\lambda\in\{3,4,6,8\}$.}
\label{fig:p0-bounded-low-lambda}
\end{figure}

When $\lambda>\lambda_{\mathrm{th}}(\delta)$ the ground state is bound,
$E_0=-\kappa_0^2<0$, described by~\eqref{eq:ring-bound-eigenfunction}. Writing it as
$\Psi_0(x) = N_0[A\sinh(\kappa_0 x)+iB\cosh(\kappa_0 x)]$ with
$A = -\sin(\delta/2)\cosh\kappa_0$ and $B = \cos(\delta/2)\sinh\kappa_0$, the same
calculation gives $\langle\hat{p}_0\rangle = -2N_0^2\kappa_0 AB$ with
$AB=-\tfrac{1}{4}\sin\delta\,\sinh(2\kappa_0)$, so again an explicit factor of
$\sin\delta$ appears. Normalising with
$N_0^{-2} = A^2 I_{\sinh^2}+B^2 I_{\cosh^2}$, where
$I_{\sinh^2}=\sinh(2\kappa_0)/(2\kappa_0)-1$ and
$I_{\cosh^2}=\sinh(2\kappa_0)/(2\kappa_0)+1$, yields
\begin{equation}
\begin{split}
\langle\hat{p}_0\rangle &= I_c^{(D)}(\kappa_0;\delta)\,\sin\delta,\\
I_c^{(D)} &:= \frac{\kappa_0\,\sinh(2\kappa_0)/2}{A^2 I_{\sinh^2}+B^2 I_{\cosh^2}},
\end{split}
\label{eq:ring-Ic-bound}
\end{equation}
with $\kappa_0=\kappa_0(\delta,\lambda)$ the positive root
of~\eqref{eq:ring-bound-secular}. In the strong-binding limit $\lambda\gg 1$ one has
$\kappa_0\approx\lambda/2$, and, keeping the leading order
for large $\kappa_0$, one gets 
\begin{equation}
I_c^{(D)} \approx 4\kappa_0^2\,e^{-2\kappa_0} \approx \lambda^2\,e^{-\lambda},
\qquad \lambda\gg 1.
\label{eq:ring-Ic-bound-limit}
\end{equation}
The sinusoidal Josephson law is thus also recovered in this case, but its amplitude is
\emph{exponentially suppressed}: a deeply bound state concentrates near the junction
and yields a small $\langle\hat{p}_0\rangle$. The leading order is compared with the
numerical results in Figure~\ref{fig:p0-bounded-low-lambda}.
This is the opposite regime to the
strong-barrier oscillatory case, where $I_c\sim\pi^2/(2|\lambda|)$ decays only as a
power law; the two limits together show how the junction controls
$\langle\hat{p}_0\rangle$ over the full $\lambda$ axis.

\paragraph{Regime summary.}
The Josephson form $\langle\hat{p}_0\rangle\approx I_c(\lambda)\sin\delta$ is
a good approximation in the strong-coupling regime $|\lambda|\gg 1$, while
noticeable corrections appear at moderate and small $|\lambda|$. The free-ring limit
is the clearest departure: at $\lambda=0$ one finds
$\langle\hat{p}_0\rangle=\delta/2$, a \emph{linear} dependence on the Josephson
phase set by the transparent junction, rather than a $\sin\delta$ Josephson law. The
model therefore predicts corrections to the first Josephson relation in the
moderate-coupling regime; \emph{these corrections should be studied and compared with
experimental data}.

\section{Time-dependent phase and the second Josephson relation}
\label{sec:second_josephson_relation}


So far the Josephson phase $\delta$ has been a fixed parameter. Let us now allow it to
vary in time, $\delta=\delta(t)$, with $\delta$ continuously differentiable, while the
coupling strength $\lambda$ is kept fixed. The Hamiltonian $H(t)$ is the Laplacian
$-\mathrm{d}^2/\mathrm{d}x^2$ on the loop $x\in(-\ell,\ell)$ with the Josephson junction
boundary conditions~\eqref{eq:JJ-BC} at phase $\delta(t)$, so that its domain changes in
time, and the dynamics is governed by the time-dependent Schr\"odinger equation
\begin{equation}
i\,\partial_t\psi(t) = H(t)\,\psi(t).
\label{eq:tdse}
\end{equation}
This problem is well posed~\cite{balmaseda2022,balmaseda2024,BalmasedaLonigroPerezPardo2023}.

As in Section~\ref{sec:planar_rotor}, the phase jump can be removed with the gauge
transformation~\eqref{eq:gauge-transformation}, which now depends on time through
$\delta(t)$. In the static case $\mathcal{U}_\delta$ maps the Hamiltonian with Josephson
junction boundary conditions onto the magnetic Hamiltonian~\eqref{eq:magnetic-laplacian};
when $\delta$ varies in time the two descriptions are no longer unitarily equivalent,
because $\mathcal{U}_{\delta(t)}$ does not commute with the time derivative. Setting
$\phi(t)=\mathcal{U}_{\delta(t)}\psi(t)$, Eq.~\eqref{eq:tdse} becomes
$i\,\partial_t\phi = \mathcal{U}_{\delta(t)}H(t)\,\mathcal{U}_{\delta(t)}^{-1}\phi
+ i\,\bigl(\partial_t\mathcal{U}_{\delta(t)}\bigr)\,\mathcal{U}_{\delta(t)}^{-1}\phi$,
that is,
\begin{equation}
i\,\partial_t\phi=\Bigl[\Bigl(-i\frac{\partial}{\partial x}+\frac{\delta(t)}{2\ell}\Bigr)^{2}
+\frac{\dot\delta(t)}{2\ell}\,x\Bigr]\phi,
\label{eq:tdse-gauge}
\end{equation}
where $\phi(t)$ now satisfies the time-independent boundary
conditions~\eqref{eq:periodic-delta-BC}. That is, besides the time-dependent vector potential
$A(t)=-\delta(t)/(2\ell)$, the time-dependent gauge transformation generates the scalar
potential
\begin{equation}
V(x,t)=\frac{\dot\delta(t)}{2\ell}\,x, \qquad x\in(-\ell,\ell),
\label{eq:induced-potential}
\end{equation}
corresponding to a uniform electric field along the loop. This is the one-dimensional
instance of the equivalence, established
in~\cite[Thm.~5.8]{BalmasedaLonigroPerezPardo2023}, between systems whose boundary
conditions carry a time-dependent phase and quantum induction systems, and of the
quantum Faraday law obtained for the planar rotor in~\cite[Example~3.5]{perezpardo2015}:
a magnetic flux that varies in time is accompanied by an electric potential proportional
to its rate of change.

The induced potential has a transparent physical meaning. 
The potential $V$ is discontinuous at the junction, where the endpoints $x=\pm\ell$ are
identified. The whole potential drop is concentrated at the junction,
\begin{equation}
V(\ell,t)-V(-\ell,t)=\dot\delta(t),
\label{eq:junction-voltage}
\end{equation}
and coincides with the electromotive force induced around the loop by the varying flux
$\Phi_B=-\delta$ (cf.\ Section~\ref{sec:planar_rotor}), $-\mathrm{d}\Phi_B/\mathrm{d}t=\dot\delta$.
Restoring units, the potential energy in~\eqref{eq:tdse-gauge} reads
$\hbar\,\dot\delta\,x/(2\ell)$, so that the voltage across the junction for carriers of
charge $q$ is $V_J=(\hbar/q)\,\mathrm{d}\delta/\mathrm{d}t$. For Cooper pairs, $q=2e$,
this is precisely the second Josephson relation~\eqref{eq:josephson-voltage}. Hence the
second Josephson relation is not an additional ingredient of the model: it emerges from
the boundary conditions~\eqref{eq:JJ-BC} once the phase is allowed to depend on time,
through the gauge transformation relating the boundary-condition and magnetic
descriptions.

\section{Conclusions}
\label{sec:conclusions}


We have proposed a purely quantum-mechanical model of a Josephson junction in which the
junction is not described by a phenomenological potential or a lumped circuit element,
but by self-adjoint boundary conditions at a point of a one-dimensional loop. The
junction is characterised by two parameters: a coupling strength $\lambda$, which governs
the jump of the derivative and hence the transparency of the contact, and a phase jump
$\delta$, which plays the role of the superconducting phase difference. A gauge
transformation shows that the phase jump is equivalent to a magnetic flux threading the
loop, so that for $\lambda=0$ the model reduces to the planar rotor in a magnetic field.

On a ring carrying a single junction, the spectral problem reduces to the scalar secular
equation~\eqref{eq:ring-scalar-secular}, which we solved in closed form along the axis
$\lambda=0$ and numerically elsewhere. The central result is that the
ground-state momentum expectation factorises exactly as
$\langle\hat{p}_0\rangle = I_c(\delta,\lambda)\sin\delta$. The $\sin\delta$ factor is not
imposed as a constitutive law; it emerges from the structure of the ground-state
eigenfunction. In the opaque limit $\lambda\to-\infty$ the coefficient becomes
independent of the phase, $I_c\simeq\pi^2/(2|\lambda|)$, and the first Josephson relation
is recovered. For strongly attractive coupling, $\lambda\gg1$, the ground state is bound
to the junction and the $\sin\delta$ law is also recovered, but with an exponentially
suppressed amplitude, $I_c\simeq\lambda^2e^{-\lambda}$.

The most distinctive outcome of the model is the \emph{prediction of deviations from the
pure $\sin\delta$ law}. Because $I_c$ depends on $\delta$ through the ground-state wave
number, the current--phase relation is not purely sinusoidal away from the limits above.
In the strong-barrier regime this can be made explicit: the $\delta$-dependent term of
order $\lambda^{-2}$ in Eq.~\eqref{eq:ring-Ic-series}, $J_1(\delta)\propto 1+\cos\delta$,
turns the current into
\begin{equation*}
\langle\hat{p}_0\rangle \simeq \Bigl(\frac{\pi^2}{2|\lambda|}-\frac{3\pi^2}{2\lambda^2}\Bigr)\sin\delta
-\frac{3\pi^2}{4\lambda^2}\,\sin 2\delta + O(\lambda^{-3}),
\end{equation*}
so that the leading deviation is a second Josephson harmonic, whose amplitude relative to
the fundamental is of order $3/(2|\lambda|)$. As the junction becomes more transparent, higher
harmonics accumulate into a sawtooth profile, and for a fully transparent contact the
current becomes linear in $\delta$, $\langle\hat{p}_0\rangle=\delta/2$. Recent spectroscopy
of superconducting qubits has shown that effective circuit models must incorporate
Josephson harmonics to reproduce measured spectra~\cite{willsch2024,kim2026}; in the
present framework such harmonics are not added phenomenologically but follow from the boundary
conditions, and they grow with the transparency of the junction, in qualitative agreement
with the interpretation of~\cite{willsch2024} in terms of high-transparency conduction
channels. Mapping $\lambda$ onto measurable junction properties, for instance through the
transmission probability, would turn these predictions into a
quantitative experimental test.

The same framework also accounts for the second Josephson relation. When the phase varies
in time, the gauge transformation that removes the phase jump generates an electric
potential whose entire drop is concentrated at the junction and equals $\hbar\dot\delta/q$;
for Cooper pairs this is precisely $V=(\hbar/2e)\,\mathrm{d}\delta/\mathrm{d}t$. Both
Josephson relations are therefore obtained from a single quantum-mechanical model,
without passing through a classical circuit.

Several directions follow naturally. Beyond the comparison with experiment, the model can
be extended to circuits with several junctions and external fields, and to networks of
higher-dimensional manifolds, where the junction becomes an interface rather than a point
and the gauge equivalence established here still holds~\cite{BalmasedaLonigroPerezPardo2023}.
A further level of refinement would incorporate the interactions between condensed
particles through the nonlinear Schr\"odinger, or Gross--Pitaevskii, equation, of which
the present Hamiltonian is the non-interacting, or low-density, limit. Finally, transport could be studied
beyond the ground-state momentum expectation, through excited states, finite temperature
or driven phases, with the aim of building complete qubit Hamiltonians directly from the
boundary-condition description and assessing their fidelity against experiment.

\section*{Acknowledgements}
JMPP acknowledges support from Agencia Estatal de Investigación, 
Project PID2024-160539NB-I00 and Comunidad de Madrid, 
Consejería de Educación, Ciencia y Universidades, Project QUITEMAD TEC-2024/COM-84.

\section*{Author contributions and use of AI}
Both authors contributed equally to the preparation of this article.

We disclose that large language models were used during the preparation of this
manuscript. They assisted with the production of the \LaTeX{} source, with language
editing and the drafting of text from the authors' instructions, with the search for
references, with analytical computations, and with the preparation of the figures.
The model, the scientific ideas and the results originate from the authors, who
verified all calculations, reviewed and edited all AI-assisted output, and take full
responsibility for the content of this work.

\bibliographystyle{unsrtnat}
\bibliography{referencias}

\begin{thebibliography}{43}
\providecommand{\natexlab}[1]{#1}
\providecommand{\url}[1]{\texttt{#1}}
\expandafter\ifx\csname urlstyle\endcsname\relax
  \providecommand{\doi}[1]{doi: #1}\else
  \providecommand{\doi}{doi: \begingroup \urlstyle{rm}\Url}\fi

\bibitem[Devoret and Schoelkopf(2013)]{devoret2013}
M.~H. Devoret and R.~J. Schoelkopf.
\newblock Superconducting circuits for quantum information: An outlook.
\newblock \emph{Science}, 339\penalty0 (6124):\penalty0 1169--1174, 2013.
\newblock \doi{10.1126/science.1231930}.

\bibitem[Jiang et~al.(2025)Jiang, Deng, Fan, et~al.]{jiang2025}
Yao-Yao Jiang, Chunqing Deng, Heng Fan, et~al.
\newblock Advancements in superconducting quantum computing.
\newblock \emph{Natl. Sci. Rev.}, 12\penalty0 (8):\penalty0 nwaf246, 2025.
\newblock \doi{10.1093/nsr/nwaf246}.

\bibitem[Bland et~al.(2025)Bland, Bahrami, Martinez, et~al.]{bland2025}
Matthew~P. Bland, Faranak Bahrami, Jeronimo G.~C. Martinez, et~al.
\newblock Millisecond lifetimes and coherence times in {2D} transmon qubits.
\newblock \emph{Nature}, 647\penalty0 (8089):\penalty0 343--348, 2025.
\newblock \doi{10.1038/s41586-025-09687-4}.

\bibitem[Nakamura et~al.(1999)Nakamura, Pashkin, and Tsai]{nakamura1999}
Y.~Nakamura, Yu.~A. Pashkin, and J.~S. Tsai.
\newblock Coherent control of macroscopic quantum states in a
  single-{Cooper}-pair box.
\newblock \emph{Nature}, 398:\penalty0 786--788, 1999.
\newblock \doi{10.1038/19718}.

\bibitem[Kim et~al.(2023)Kim, Eddins, Anand, et~al.]{kim2023}
Youngseok Kim, Andrew Eddins, Sajant Anand, et~al.
\newblock Evidence for the utility of quantum computing before fault tolerance.
\newblock \emph{Nature}, 618\penalty0 (7965):\penalty0 500--505, 2023.
\newblock \doi{10.1038/s41586-023-06096-3}.

\bibitem[{Google Quantum AI and Collaborators} et~al.(2025){Google Quantum AI
  and Collaborators}, Acharya, Abanin, et~al.]{google2025}
{Google Quantum AI and Collaborators}, Rajeev Acharya, Dmitry~A. Abanin, et~al.
\newblock Quantum error correction below the surface code threshold.
\newblock \emph{Nature}, 638\penalty0 (8052):\penalty0 920--926, 2025.
\newblock \doi{10.1038/s41586-024-08449-y}.

\bibitem[Gao et~al.(2025)Gao, Fan, Zha, et~al.]{gao2025}
Dongxin Gao, Daojin Fan, Chen Zha, et~al.
\newblock Establishing a new benchmark in quantum computational advantage with
  105-qubit {Zuchongzhi} 3.0 processor.
\newblock \emph{Phys. Rev. Lett.}, 134\penalty0 (9):\penalty0 090601, 2025.
\newblock \doi{10.1103/PhysRevLett.134.090601}.

\bibitem[Josephson(1962)]{josephson1962}
B.~D. Josephson.
\newblock Possible new effects in superconductive tunnelling.
\newblock \emph{Phys. Lett.}, 1\penalty0 (7):\penalty0 251--253, 1962.
\newblock \doi{10.1016/0031-9163(62)91369-0}.

\bibitem[Blais et~al.(2021)Blais, Grimsmo, Girvin, and Wallraff]{blais2021}
A.~Blais, A.~L. Grimsmo, S.~M. Girvin, and A.~Wallraff.
\newblock Circuit quantum electrodynamics.
\newblock \emph{Rev. Mod. Phys.}, 93:\penalty0 025005, 2021.
\newblock \doi{10.1103/RevModPhys.93.025005}.

\bibitem[Wendin and Shumeiko(2007)]{wendin2007}
G.~Wendin and V.~S. Shumeiko.
\newblock Quantum bits with {Josephson} junctions.
\newblock \emph{Low Temp. Phys.}, 33\penalty0 (9):\penalty0 724--744, 2007.
\newblock \doi{10.1063/1.2780165}.

\bibitem[Wendin(2017)]{wendin2017}
G.~Wendin.
\newblock Quantum information processing with superconducting circuits: a
  review.
\newblock \emph{Rep. Prog. Phys.}, 80\penalty0 (10):\penalty0 106001, 2017.
\newblock \doi{10.1088/1361-6633/aa7e1a}.

\bibitem[Koch et~al.(2007)Koch, Yu, Gambetta, Houck, Schuster, Majer, Blais,
  Devoret, Girvin, and Schoelkopf]{koch2007}
J.~Koch, T.~M. Yu, J.~Gambetta, A.~A. Houck, D.~I. Schuster, J.~Majer,
  A.~Blais, M.~H. Devoret, S.~M. Girvin, and R.~J. Schoelkopf.
\newblock Charge-insensitive qubit design derived from the {Cooper} pair box.
\newblock \emph{Phys. Rev. A}, 76:\penalty0 042319, 2007.
\newblock \doi{10.1103/PhysRevA.76.042319}.

\bibitem[Smith et~al.(2025)Smith, Leu, Miyanishi, Gely, and Lucas]{smith2025}
M.~C. Smith, A.~D. Leu, K.~Miyanishi, M.~F. Gely, and D.~M. Lucas.
\newblock Single-qubit gates with errors at the $10^{-7}$ level.
\newblock \emph{Phys. Rev. Lett.}, 134\penalty0 (23):\penalty0 230601, 2025.
\newblock \doi{10.1103/42w2-6ccy}.

\bibitem[Ransford et~al.(2026)Ransford, Allman, Arkinstall,
  et~al.]{ransford2026}
Anthony Ransford, M.~S. Allman, Jake Arkinstall, et~al.
\newblock A 98-qubit trapped-ion quantum computer with all-to-all connectivity.
\newblock \emph{Nature}, 655\penalty0 (8121):\penalty0 81--86, 2026.
\newblock \doi{10.1038/s41586-026-10676-4}.

\bibitem[Nigg et~al.(2012)Nigg, Paik, Vlastakis, Kirchmair, Shankar, Frunzio,
  Devoret, Schoelkopf, and Girvin]{nigg2012}
Simon~E. Nigg, Hanhee Paik, Brian Vlastakis, Gerhard Kirchmair, S.~Shankar,
  Luigi Frunzio, M.~H. Devoret, R.~J. Schoelkopf, and S.~M. Girvin.
\newblock Black-box superconducting circuit quantization.
\newblock \emph{Phys. Rev. Lett.}, 108\penalty0 (24):\penalty0 240502, 2012.
\newblock \doi{10.1103/PhysRevLett.108.240502}.

\bibitem[Minev et~al.(2021)Minev, Leghtas, Mundhada, Christakis, Pop, and
  Devoret]{minev2021}
Zlatko~K. Minev, Zaki Leghtas, Shantanu~O. Mundhada, Lysander Christakis,
  Ioan~M. Pop, and Michel~H. Devoret.
\newblock Energy-participation quantization of {Josephson} circuits.
\newblock \emph{npj Quantum Inf.}, 7\penalty0 (1):\penalty0 131, 2021.
\newblock \doi{10.1038/s41534-021-00461-8}.

\bibitem[Yilmaz et~al.(2026)Yilmaz, Singh, Zwanenburg, Hu, Stefanski, and
  Andersen]{yilmaz2026}
Figen Yilmaz, Siddharth Singh, Martijn F.~S. Zwanenburg, Jinlun Hu, Taryn~V.
  Stefanski, and Christian~Kraglund Andersen.
\newblock Energy-participation-ratio analysis for very anharmonic
  superconducting circuits.
\newblock \emph{Phys. Rev. Appl.}, 25\penalty0 (4):\penalty0 044021, 2026.
\newblock \doi{10.1103/1rbn-c4xf}.

\bibitem[Ciani et~al.(2024)Ciani, DiVincenzo, and Terhal]{ciani2024}
Alessandro Ciani, David~P. DiVincenzo, and Barbara~M. Terhal.
\newblock \emph{Lecture Notes on Quantum Electrical Circuits}.
\newblock TU Delft OPEN Publishing, Delft, 2024.
\newblock \doi{10.59490/tb.85}.

\bibitem[Rasmussen et~al.(2021)Rasmussen, Christensen, Pedersen, Kristensen,
  B{\ae}kkegaard, Loft, and Zinner]{rasmussen2021}
S.~E. Rasmussen, K.~S. Christensen, S.~P. Pedersen, L.~B. Kristensen,
  T.~B{\ae}kkegaard, N.~J.~S. Loft, and N.~T. Zinner.
\newblock Superconducting circuit companion---an introduction with worked
  examples.
\newblock \emph{PRX Quantum}, 2\penalty0 (4):\penalty0 040204, 2021.
\newblock \doi{10.1103/PRXQuantum.2.040204}.

\bibitem[Osborne et~al.(2024)Osborne, Larson, Jones, Simmonds, Gyenis, and
  Lucas]{osborne2024}
Andrew Osborne, Trevyn Larson, Sarah~Garcia Jones, Ray~W. Simmonds, Andr\'as
  Gyenis, and Andrew Lucas.
\newblock Symplectic geometry and circuit quantization.
\newblock \emph{PRX Quantum}, 5\penalty0 (2):\penalty0 020309, 2024.
\newblock \doi{10.1103/PRXQuantum.5.020309}.

\bibitem[Parra-Rodriguez and Egusquiza(2024)]{parrarodriguez2024}
A.~Parra-Rodriguez and I.~L. Egusquiza.
\newblock Geometrical description and {Faddeev-Jackiw} quantization of
  electrical networks.
\newblock \emph{Quantum}, 8:\penalty0 1466, 2024.
\newblock \doi{10.22331/q-2024-09-09-1466}.

\bibitem[Park et~al.(2026)Park, Choi, Kim, Park, Choi, Choi, Chong, Lee, and
  Hahn]{park2024accuracy}
Seong~Hyeon Park, Gahyun Choi, Eunjong Kim, Gwanyeol Park, Jisoo Choi, Jiman
  Choi, Yonuk Chong, Yong-Ho Lee, and Seungyong Hahn.
\newblock Kinetic-inductance-incorporated quantization for accurate
  {Hamiltonian} prediction in superconducting circuits.
\newblock \emph{npj Quantum Inf.}, 2026.
\newblock \doi{10.1038/s41534-026-01187-1}.
\newblock arXiv:2410.24004.

\bibitem[Willsch et~al.(2024)Willsch, Rieger, Winkel, et~al.]{willsch2024}
Dennis Willsch, Dennis Rieger, Patrick Winkel, et~al.
\newblock Observation of {Josephson} harmonics in tunnel junctions.
\newblock \emph{Nat. Phys.}, 20\penalty0 (5):\penalty0 815--821, 2024.
\newblock \doi{10.1038/s41567-024-02400-8}.

\bibitem[Kim et~al.(2026)Kim, Hays, Rosen, et~al.]{kim2026}
Junghyun Kim, Max Hays, Ilan~T. Rosen, et~al.
\newblock Higher-order harmonics in {Josephson} tunnel junctions due to series
  inductance.
\newblock \emph{Nat. Phys.}, 22\penalty0 (6):\penalty0 897--902, 2026.
\newblock \doi{10.1038/s41567-026-03285-5}.

\bibitem[Liao et~al.(2025)Liao, Powell, and Stace]{liao2025}
Yun-Chih Liao, Ben~J. Powell, and Thomas~M. Stace.
\newblock Circuit quantization from first principles.
\newblock \emph{Phys. Rev. Res.}, 7\penalty0 (3):\penalty0 033144, 2025.
\newblock \doi{10.1103/dfrq-44vk}.

\bibitem[von Neumann(1932)]{vonneumann1932}
J.~von Neumann.
\newblock \emph{Mathematische {Grundlagen} der {Quantenmechanik}}.
\newblock Springer, Berlin, 1932.
\newblock URL \url{https://link.springer.com/book/10.1007/978-3-642-96048-2}.

\bibitem[Grubb(1968)]{grubb1968}
G.~Grubb.
\newblock A characterization of the nonlocal boundary value problems associated
  with an elliptic operator.
\newblock \emph{Ann. Sc. Norm. Super. Pisa}, 22\penalty0 (3):\penalty0
  425--513, 1968.
\newblock URL \url{http://www.numdam.org/item/ASNSP_1968_3_22_3_425_0/}.

\bibitem[Kochubei(1975)]{kochubei1975}
A.~N. Kochubei.
\newblock Extensions of symmetric operators and symmetric binary relations.
\newblock \emph{Math. Notes}, 17\penalty0 (1):\penalty0 25--28, 1975.
\newblock \doi{10.1007/BF01093837}.
\newblock Translated from Matematicheskie Zametki.

\bibitem[Asorey et~al.(2005)Asorey, Ibort, and Marmo]{asorey2005}
M.~Asorey, A.~Ibort, and G.~Marmo.
\newblock Global theory of quantum boundary conditions and topology change.
\newblock \emph{Int. J. Mod. Phys. A}, 20\penalty0 (05):\penalty0 1001--1025,
  2005.
\newblock \doi{10.1142/S0217751X05019798}.

\bibitem[Ibort et~al.(2015{\natexlab{a}})Ibort, Lled\'o, and
  P\'erez-Pardo]{ibortlledojfa2015}
A.~Ibort, F.~Lled\'o, and J.~M. P\'erez-Pardo.
\newblock Self-adjoint extensions of the {Laplace-Beltrami} operator and
  unitaries at the boundary.
\newblock \emph{J. Funct. Anal.}, 268:\penalty0 634--670, 2015{\natexlab{a}}.
\newblock \doi{10.1016/j.jfa.2014.10.013}.

\bibitem[Ibort and P\'erez-Pardo(2015)]{ibort2015}
A.~Ibort and J.~M. P\'erez-Pardo.
\newblock On the theory of self-adjoint extensions of symmetric operators and
  its applications to quantum physics.
\newblock \emph{Int. J. Geom. Methods Mod. Phys.}, 12\penalty0 (07):\penalty0
  1560005, 2015.
\newblock \doi{10.1142/S0219887815600051}.

\bibitem[Ibort et~al.(2015{\natexlab{b}})Ibort, Lled\'o, and
  P\'erez-Pardo]{ibortlledo2015}
A.~Ibort, F.~Lled\'o, and J.~M. P\'erez-Pardo.
\newblock On self-adjoint extensions and symmetries in quantum mechanics.
\newblock \emph{Ann. Henri Poincar\'e}, 16\penalty0 (10):\penalty0 2367--2397,
  2015{\natexlab{b}}.
\newblock \doi{10.1007/s00023-014-0379-4}.

\bibitem[Albeverio et~al.(2005)Albeverio, Gesztesy, H{\o}egh-Krohn, and
  Holden]{albeverio2005}
S.~Albeverio, F.~Gesztesy, R.~H{\o}egh-Krohn, and H.~Holden.
\newblock \emph{Solvable Models in Quantum Mechanics}.
\newblock American Mathematical Society (AMS Chelsea Publishing), Providence,
  RI, 2nd edition, 2005.
\newblock \doi{10.1090/chel/350}.

\bibitem[Br\"uning et~al.(2008)Br\"uning, Geyler, and Pankrashkin]{bruning2008}
Jochen Br\"uning, Vladimir Geyler, and Konstantin Pankrashkin.
\newblock Spectra of self-adjoint extensions and applications to solvable
  {Schr\"odinger} operators.
\newblock \emph{Rev. Math. Phys.}, 20\penalty0 (1):\penalty0 1--70, 2008.
\newblock \doi{10.1142/S0129055X08003249}.

\bibitem[Balmaseda et~al.(2019)Balmaseda, {di Cosmo}, and
  P\'erez-Pardo]{balmaseda2019}
A.~Balmaseda, F.~{di Cosmo}, and J.~M. P\'erez-Pardo.
\newblock On {$\mathbb{Z}$}-invariant self-adjoint extensions of the laplacian
  on quantum circuits.
\newblock \emph{Symmetry}, 11\penalty0 (8):\penalty0 1047, 2019.
\newblock \doi{10.3390/sym11081047}.

\bibitem[Balmaseda et~al.(2023)Balmaseda, Lonigro, and
  P{\'e}rez-Pardo]{BalmasedaLonigroPerezPardo2023}
Aitor Balmaseda, Davide Lonigro, and Juan~Manuel P{\'e}rez-Pardo.
\newblock Quantum controllability on graph-like manifolds through magnetic
  potentials and boundary conditions.
\newblock \emph{Journal of Physics A: Mathematical and Theoretical},
  56\penalty0 (32):\penalty0 325201, 2023.
\newblock \doi{10.1088/1751-8121/ace505}.

\bibitem[Angelone et~al.(2023)Angelone, Asorey, Facchi, Lonigro, and
  Martinez]{angelone2023}
Giuliano Angelone, Manuel Asorey, Paolo Facchi, Davide Lonigro, and Yisely
  Martinez.
\newblock Boundary conditions for the quantum {Hall} effect.
\newblock \emph{J. Phys. A: Math. Theor.}, 56\penalty0 (2):\penalty0 025301,
  2023.
\newblock \doi{10.1088/1751-8121/acb026}.

\bibitem[Deelan~Cunden et~al.(2026)Deelan~Cunden, Gramegna, and
  Ligab\`o]{cunden2026}
Fabio Deelan~Cunden, Giovanni Gramegna, and Marilena Ligab\`o.
\newblock Quantum systems with jump-discontinuous mass. {I}.
\newblock \emph{Nonlinearity}, 39\penalty0 (4):\penalty0 045010, 2026.
\newblock \doi{10.1088/1361-6544/ae4afb}.

\bibitem[Asorey et~al.(1983)Asorey, Esteve, and Pacheco]{asorey1983}
M.~Asorey, J.~G. Esteve, and A.~F. Pacheco.
\newblock Planar rotor: The $\theta$-vacuum structure, and some approximate
  methods in quantum mechanics.
\newblock \emph{Phys. Rev. D}, 27\penalty0 (8):\penalty0 1852--1868, 1983.
\newblock \doi{10.1103/PhysRevD.27.1852}.

\bibitem[Berkolaiko and Kuchment(2013)]{berkolaiko2013}
Gregory Berkolaiko and Peter Kuchment.
\newblock \emph{Introduction to Quantum Graphs}, volume 186 of
  \emph{Mathematical Surveys and Monographs}.
\newblock American Mathematical Society, Providence, RI, 2013.
\newblock \doi{10.1090/surv/186}.

\bibitem[Balmaseda et~al.(2022)Balmaseda, Lonigro, and
  P\'erez-Pardo]{balmaseda2022}
Aitor Balmaseda, Davide Lonigro, and Juan~Manuel P\'erez-Pardo.
\newblock On the {Schr\"odinger} equation for time-dependent {Hamiltonians}
  with a constant form domain.
\newblock \emph{Mathematics}, 10\penalty0 (2):\penalty0 218, 2022.
\newblock \doi{10.3390/math10020218}.

\bibitem[Balmaseda et~al.(2024)Balmaseda, Lonigro, and
  P\'erez-Pardo]{balmaseda2024}
Aitor Balmaseda, Davide Lonigro, and Juan~Manuel P\'erez-Pardo.
\newblock On a sharper bound on the stability of non-autonomous {Schr\"odinger}
  equations and applications to quantum control.
\newblock \emph{J. Funct. Anal.}, 287\penalty0 (8):\penalty0 110563, 2024.
\newblock \doi{10.1016/j.jfa.2024.110563}.

\bibitem[P\'erez-Pardo et~al.(2015)P\'erez-Pardo, Barbero-Li\~n\'an, and
  Ibort]{perezpardo2015}
J.~M. P\'erez-Pardo, M.~Barbero-Li\~n\'an, and A.~Ibort.
\newblock Boundary dynamics and topology change in quantum mechanics.
\newblock \emph{Int. J. Geom. Methods Mod. Phys.}, 12\penalty0 (8):\penalty0
  1560011, 2015.
\newblock \doi{10.1142/S0219887815600117}.

\end{thebibliography}

\end{document}